\documentclass[11pt,a4paper]{article}

\usepackage{jheppub}
\usepackage{chessfss}

\newcommand{\beqa}{\begin{eqnarray}}
\newcommand{\eeqa}{\end{eqnarray}}

\newcommand{\beq}{\begin{equation}}
\newcommand{\eeq}{\end{equation}}
\newcommand{\bea}{\begin{eqnarray}}
\newcommand{\eea}{\end{eqnarray}}

\newcommand{\CB}{{\mathcal B}}
\newcommand{\CC}{{\mathcal C}}

\newcommand{\CK}{{\mathcal K}}

\newcommand{\CM}{{\mathcal M}}
\newcommand{\CN}{{\mathcal N}}

\newcommand{\CQ}{{\mathcal Q}}

\newcommand\SU{{\rm SU}}
\newcommand\U{{\rm U}}

\def\Tr{\mathop{\rm Tr}}
\newcommand\tr{\mathrm{tr}}

\newcommand\diag{\mathrm{diag}}

\newcommand{\bfa}{{\bf a}}
\newcommand{\bfb}{{\bf b}}
\newcommand{\bfc}{{\bf c}}

\def\tr{\mathop{\mathrm{tr}}\nolimits}

\def\SU{\mathrm{SU}}

\def\cN{\mathcal{N}}

\def\vev#1{\langle#1\rangle}

\def\SU{\mathrm{SU}}

\def\cI{\mathcal{I}}
\def\cN{\mathcal{N}}

\def\diag{\mathop{\mathrm{diag}}}
\def\tr{\mathop{\mathrm{tr}}\nolimits}
\def\vev#1{\langle#1\rangle}

\def\e{\epsilon}

\def\CB{{\cal B}}
\def\CC{{\cal C}}

\def\CK{{\cal K}}

\def\CM{{\cal M}}
\def\CN{{\cal N}}

\def\CQ{{\cal Q}}

\def\diag{\mathop{\rm diag}\nolimits}

\def\SU{\mathop{\rm SU}}
\let\U\UU

\def\tr{\mathop{\rm tr}}

\def\beq#1\eeq{\begin{align}#1\end{align}}

\newcommand{\beqr}{\begin{array}}
\newcommand {\eeqr}{\end{array}}

\setboardfontsize{8}

\title{$\CN{=}1$ dynamics with $T_N$ theory
}

\preprint{CALT-68-2936, IPMU-13-0099, UT-13-20}

\author[\WhiteKnightOnWhite]{Kazunobu Maruyoshi,}
\affiliation[\WhiteKnightOnWhite]{California Institute of Technology \\ Pasadena, CA 91125, USA}
\emailAdd{maruyosh@caltech.edu}

\author[\WhitePawnOnWhite]{Yuji Tachikawa,}
\affiliation[\WhitePawnOnWhite]{Department of Physics, Faculty of Science, \\
 University of Tokyo,  Bunkyo-ku, Tokyo 133-0022, Japan and \\
Institute for the Physics and Mathematics of the Universe (WPI), \\
 University of Tokyo,  Kashiwa, Chiba 277-8583, Japan}
\emailAdd{yuji.tachikawa@ipmu.jp}

\author[\WhiteKnightOnWhite]{Wenbin Yan,}
\emailAdd{wbyan@theory.caltech.edu}

\author[\WhiteKingOnWhite]{and Kazuya Yonekura}
\affiliation[\WhiteKingOnWhite]{School of Natural Sciences, Institute for Advanced Study,\\
1 Einstein Drive, Princeton, NJ 08540 USA }
\emailAdd{yonekura@ias.edu}

\abstract{
We study the dynamics of $\cN=1$ supersymmetric systems consisting of   the strongly-coupled superconformal theory $T_N$, $\SU(N)$ gauge groups, and fundamental chiral multiplets.
We demonstrate that such systems exhibit familiar phenomena such as deformation of the vacuum moduli space, appearance of the dynamical superpotential, and Coulomb branches with $\cN=1$ Seiberg-Witten curves.  The analysis requires a rather detailed knowledge of the chiral ring of the $T_N$ theory, which will also be discussed at length.

}


\begin{document}
\setcounter{tocdepth}{2}
\maketitle
\section{Introduction and summary}
\label{sec:intro}

Since the seminal work \cite{Seiberg:1994bz,Seiberg:1994pq}, people have extensively studied the dynamics of $\cN=1$ conventional supersymmetric QFTs, made of gauge multiplets and elementary chiral multiplets.
It is becoming more and more apparent, however, that we also need to study the dynamics of non-conventional supersymmetric QFTs, made of gauge multiplets coupled to already strongly-coupled supersymmetric theories, in order to fully understand the duality. This is because the dual of a conventional theory can often be non-conventional \cite{Argyres:2007cn,Gaiotto:2009we,Gadde:2013fma}.

So far the study of the non-conventional theories was restricted to the case when they flow to nontrivial superconformal theories in the infrared. In this paper we study non-conventional theories in a different regime, where they flow to almost free theories. The basic ingredient is the $T_N$ theory introduced by \cite{Gaiotto:2009we,Gaiotto:2009gz}, which is an $\cN=2$ supersymmetric strongly-coupled theory with $\SU(N)^3$ flavor symmetry.
If we gauge one of three $\SU(N)$ symmetries, the contribution to the one-loop beta function is the same as that from $N$ flavors of chiral multiplets in the fundamental and the antifundamental representations.
By giving a suitable vev, the $T_N$ theory flows to a system of free bifundamental chiral multiplets.
We distinguish three $\SU(N)$ symmetries by calling them $\SU(N)_X$ for $X=A,B,C$.
The $T_N$ theory has dimension-2 chiral operators $\mu_X$, which transforms as adjoints of $\SU(N)_X$, for each $X=A,B,C$. They satisfy the constraints \begin{equation}
\tr \mu_A^k = \tr \mu_B^k = \tr \mu_C^k,\label{oconstraint}
\end{equation} or equivalently \begin{equation}
 v_{A,k} =  v_{B,k} = v_{C,k}, \quad \text{where} \quad \det(x-\mu_X)= x^N + v_{X,2} x^{N-2} + \cdots + v_{X,N}.
\end{equation}
This theory also has operators $Q^{i_A i_B i_C}$ and $Q_{i_Ai_Bi_C}$ of dimension $N-1$, transforming as trifundamentals under three $\SU(N)$ symmetries.

We will study the following three cases in detail: \begin{itemize}
\item The $T_N$ theory coupled to one $\cN=1$ $\SU(N)$ gauge multiplet. This leads to the quantum deformation of the vacuum moduli space of the $T_N$ theory. For definiteness let us gauge $\SU(N)_C$. Then the deformed moduli space is given by deforming the constraints \eqref{oconstraint} to \begin{equation}
\tr \mu_A^k - \tr \mu_B^k = \Lambda^{2N} \delta^{N,k},
\end{equation}  where $\Lambda$ is the dynamical scale of the gauge theory.
We can give a suitable vev and make the system reproduce the deformed moduli space of $\SU(N)$ theory with $N$ flavors.
\item The system above plus one flavor of fundamental  and antifundamental chiral multiplets, $q^{i_C}$ and $\tilde q_{i_C}$.
Here, the IR description involves gauge-invariant composite fields behaving almost freely, with a dynamically-generated superpotential. In more detail, the list of the gauge invariant operators of this theory is
    \beq
    &{B}_{i_A i_B}
     =     Q_{i_A i_B i_C}q^{i_C}, ~~~
    \tilde{B}^{i_A i_B}
     =     Q^{i_A i_B i_C} \tilde{q}_{i_C}, \nonumber \\
    &\mu_X~~(X=A,B), ~~~
    {M}_k
     =     \tilde{q} \mu_C^k q~~(k=0,1,\cdots,N-2).
    \eeq
Then the non-perturbative superpotential is
    \beq
    W
     =    - \frac{1}{\Lambda^{2N-1}}\left[\sum_{k=0}^{N-2} {M}_k f_{N-k}(\mu_A,\mu_B) -\tr [({B}\tilde{B})_A\mu_A]
         + \tr [(\tilde{B} { B})_B \mu_B] \right],
    \eeq
  where $f_{N-k}(\mu_A,\mu_B)$  is given by
    \beq
    f_{N-k}(\mu_A,\mu_B)
     =     \sum_{\ell=0}^{N-k-2} \frac{ v_{A,\ell}}{N-k-\ell} (\tr \mu_A^{N-k-\ell} - \tr \mu_B^{N-k-\ell} ).
    \eeq
 Again, by giving a vev, the system reproduces the dynamics of $\SU(N)$ theory with $N+1$ flavors.

\item The $T_N$ theory coupled to three $\SU(N)$ gauge multiplets.  The IR theory on a generic point on the vacuum moduli space is given by $2(N-1)$ almost-free $\U(1)$ vector multiplets and $(N-1)$ neutral chiral multiplets $v_{2,\ldots,N}$. We determine the $\cN=1$ Seiberg-Witten curve describing the holomorphic coupling constant matrices. It has the form
   \beq
    w^{N} + v_2 w^{N-2} + \cdots + v_{N-1} w + v_N  - t \Lambda^{2N}_A-\frac{\Lambda^{2N}_B}{t} - \frac{\Lambda^{2N}_C}{t-1} =0.
    \eeq
 We also study the dynamics of a class of ${\cal N}=2$ theories specified by
 a Riemann surface with several full punctures~\cite{Gaiotto:2009we},
 when their flavor symmetries are gauged by ${\cal N}=1$ vector multiplets.
 It is also possible to add some flavors of quarks to these gauge groups.
 We propose that the curve of these systems are given by
 \beq
 w^{N}+\sum_{k=2}^N V_k(t)w^{N-k}=0,
 \eeq
 where $V_k(t)$ are meromorphic functions on the Riemann surface with simple poles at the punctures.
 This is a generalization to $N>2$ of the result of \cite{Tachikawa:2011ea}.
\end{itemize}

The techniques we use in the analysis are holomorphy, matching of 't Hooft anomalies, the behavior under decoupling of the flavors, and so on. They are the same ones we use to analyze conventional $\cN=1$ supersymmetric gauge theories; the point is that these standard methods do not  rely on the existence of a Lagrangian description with gauge multiplets and chiral multiplets only, and therefore are also applicable to non-conventional theories.

Since we use the $T_N$ theory as one of the ingredients of the system we study, a rather extensive knowledge of its properties is necessary. Those properties can be deduced from other known dualities and from the known superconformal index \cite{Gadde:2011uv}, which we will discuss in detail. The overall consistency of the properties of the $T_N$ theory thus obtained and the behavior of the $T_N$ theory coupled to $\SU(N)$ gauge multiplets gives nontrivial checks of the whole procedure.

  The rest of the paper is organized as follows.
  In section \ref{sec:higgs} we study the chiral ring of the $T_{N}$ theory
  which will be extensively used in the subsequent sections.
  In section \ref{sec:deformed} the low energy behavior
  of the $T_{N}$ theory coupled to the $\CN=1$ $\SU(N)$ gauge group is proposed.
  We check this proposal by performing 't Hooft anomaly matching
  and showing that it reduces to $\CN=1$ $\SU(N)$ gauge theory with $N$ flavors by a suitable vev.
  The addition of one flavor to the previous case is explored in section~\ref{sec:Nc+1}.
  The similar checks are performed.
  In section \ref{sec:Coulomb} we see that the $T_{N}$ theory coupled to three $\CN=1$ $\SU(N)$ gauge groups
  is in the abelian Coulomb phase and derive the $\CN=1$ Seiberg-Witten curve
  encoding the holomorphic coupling constant matrix of low energy $\U(1)$ gauge groups.
  We study the case where the gauge groups are also coupled to flavors.
  Finally we generalize this to a large class of $\CN=1$ theories
  which consists of a couple of $T_{N}$ theories and gauge groups.
  We conclude with brief discussions in section \ref{sec:discussion}.
  In appendix \ref{appsec:ConstraintsClosingPunc},
  we show the detailed calculation of the reduction to $\CN=1$ gauge theories with $N$ and $N+1$ flavors discussed
  in sections \ref{sec:deformed} and \ref{sec:Nc+1}.

\section{The chiral ring of $T_N$ theory}
\label{sec:higgs}
\subsection{Summary of chiral operators}
The $T_N$ theory is an ${\cal N}=2$ SCFT which has $\SU(N)_A \times \SU(N)_B \times \SU(N)_C$ flavor symmetries.
In terms of ${\cal N}=1$ language, this theory has the following set of Higgs branch chiral operators;
\begin{enumerate}
\item Operators $\mu_A$, $\mu_B$ and $\mu_C$ transforming in the adjoint representations of $\SU(N)_A$, $\SU(N)_B$ and $\SU(N)_C$, respectively.
They are the holomorphic moment maps associated to these flavor symmetries in the hyperk\"ahler manifold. They have scaling dimension $2$.
\item Operators $Q_{(k)}~(k=1,\cdots,N-1)$ in the $(\wedge^k,\wedge^k,\wedge^k)$ representations of $\SU(N)_A \times \SU(N)_B \times \SU(N)_C$,
where $\wedge^k$ is the $k$-index anti-symmetric tensor representation (see \cite{Gaiotto:2009gz} for $k=1,N-1$ and
\cite{Gadde:2013fma} for others). 
$Q_{(k)}$ has scaling dimension $k(N-k)$. In particular, the operators with $k=1$ and $k=N-1$ are denoted as
$Q^{i_A i_B i_C}$ and $Q_{i_A i_B i_C}$ respectively, where $i_X=1,\ldots,N$ for $X=A,B,C$ are the indices of $\SU(N)_X$.
\end{enumerate}
In this paper, we do not discuss the operators $Q_{(k)}$ with $k=2,\cdots,N-2$ in detail; their presence does not affect our analysis.

In addition to the above flavor symmetries, there are $\U(1)_{R({\cal N}=2)} \times \SU(2)_R$ $R$-symmetries of the ${\cal N}=2$ superconformal algebra.
All the lowest components of $\mu_{A,B,C}$ and $Q_{(k)}$ are neutral under the $\U(1)_{R({\cal N}=2)}$
symmetry, while their charges under $\U(1)_{2I_{3}}$ are
equal to the scaling dimensions of these fields, where $\U(1)_{2I_{3}}$ is twice the Cartan sugalgebla of $\SU(2)_R$.
For example, $\mu_{A,B,C}$ have $\U(1)_{2I_{3}}$ charge $2$ and $Q_{(k)}$ has $k(N-k)$. 

We also have Coulomb branch operators $u_{d,i}$ with dimension $d$, zero $\U(1)_{2I_3}$ charge, and $\U(1)_{R(\cN=2)}$ charge $2d$, for $d=3,4,\ldots, N$ and $i=1,\ldots,d-2$: \begin{equation}
u_{3,1};\quad
u_{4,1}, u_{4,2}; \quad
u_{5,1}, u_{5,2}, u_{5,3}; \quad \ldots \quad;
u_{N,1}, u_{N,2}, \ldots, u_{N,N-2}.
\end{equation}

We discuss the chiral ring relations among these operators in detail below. Those readers who are mainly interested in the dynamics of the $\cN=1$ systems with $T_N$ theories can skip to the next section, coming back to this section only when it is necessary.

\subsection{Summary of chiral ring relations}\label{sec:CR}
First,  there are no relations among the Coulomb branch operators $u_{d,i}$. This is an analogue of the statement that there are no relations among the operators $\tr\phi^k$, $k=2,\ldots, N$, parameterizing the Coulomb branch of an $\cN=2$ $\SU(N)$ gauge theory. This feature, of the complete absence of the chiral ring relations among the Coulomb branch operators, is believed to be generic for any $\cN=2$ superconformal field theory in four dimensions.

Second, there are generic relations between Coulomb branch and Higgs branch operators of the form
\begin{equation}
u X =0
\end{equation} where $X$ is a generic Higgs branch operator and $u$ is a generic Coulomb branch operator.
It is possible that some of them satisfy more complicated relations, due to the possibility that on a sublocus on the Coulomb branch there can be a smaller Higgs branch, etc. In this paper we focus on a generic point of the Higgs branch, and hence all the Coulomb branch directions are lifted unless
otherwise stated.

The rest of the discussions is devoted to the chiral ring relations among Higgs branch operators.
Let us define the characteristic polynomial of $\mu_X~(X=A,B,C)$ as
\beq
P_X(x)=\det(x {\bf 1}-\mu_X)=\sum_{k=0}^N v_{X,k} x^{N-k}, \label{eq:charateristicP}
\eeq
where $v_{X,0}=1$ and $v_{X,1}=0$.
We claim that the chiral operators satisfy the following chiral ring relations\footnote{However, we do not claim that these are the complete
list of chiral ring relations. At least there should be relations involving $Q_{(k)}$ for $k=2,\cdots,N-2$.
};
\beq
&\tr \mu_A^k=\tr \mu_B^k=\tr \mu_C^k~~~(k=2,\cdots,N), \label{eq:CR1}
\eeq
\beq
(\mu_A)^{i_A}_{~j_A}Q^{j_A i_B i_C}=(\mu_B)^{i_B}_{~j_B}Q^{i_A j_B i_C}=(\mu_C)^{i_C}_{~j_C}Q^{i_A i_B j_C}, \label{eq:CR2} \\
(\mu_A)^{j_A}_{~i_A}Q_{j_A i_B i_C}=(\mu_B)_{~i_B}^{j_B}Q_{i_A j_B i_C}=(\mu_C)_{~i_C}^{j_C}Q_{i_A i_B j_C}, \label{eq:CR2'}
\eeq
\beq
Q^{i_A i_B i_C}Q_{j_A j_B i_C}=\sum_{l=0}^N  v_l  \sum_{m=0}^{N-l-1} (\mu_A^{N-l-1-m})^{i_A}_{~j_A} (\mu_B^m)^{i_B}_{~j_B}, \label{eq:CR3}
\eeq
\beq
&\frac{1}{(N-1)!}Q^{i_{A,1} i_{B,1} i_{C,1}}Q^{i_{A,2} i_{B,2} i_{C,2}}\cdots Q^{i_{A,N-1} i_{B,N-1} i_{C,N-1}}
\epsilon_{ i_{B,1}i_{B,2}\cdots i_{B,N-1}i_B }\epsilon_{ i_{C,1}i_{C,2}\cdots i_{C,N-1} i_C} \nonumber \\
&= Q_{i_A i_B i_C} (\mu_A^0)^{(i_{A,1}}_{~j_{A,1}} (\mu_A)^{i_{A,2}}_{~j_{A,2}} (\mu^2_A)^{i_{A,3}}_{~j_{A,3}}\cdots
(\mu^{N-2}_A)^{i_{A,N-1})}_{~j_{A,N-1}}
\epsilon^{  j_{A,1} j_{A,2} j_{A,3}\cdots   j_{A,N-1} i_A}, \label{eq:CR4}
\eeq
\beq
&\frac{1}{(N-1)!}Q_{i_{A,1} i_{B,1} i_{C,1}}Q_{i_{A,2} i_{B,2} i_{C,2}}\cdots Q_{i_{A,N-1} i_{B,N-1} i_{C,N-1}}
\epsilon^{ i_{B,1}i_{B,2}\cdots i_{B,N-1} i_B}\epsilon^{ i_{C,1}i_{C,2}\cdots i_{C,N-1} i_C} \nonumber \\
&=(-1)^{\frac{1}{2}N(N-1)} Q^{i_A i_B i_C} (\mu^0_A)^{j_{A,1}}_{~(i_{A,1}} (\mu_A)^{j_{A,2}}_{~i_{A,2}}(\mu^2_A)^{j_{A,3}}_{~i_{A,3}}
\cdots (\mu^{N-2}_A)^{j_{A,N-1}}_{~i_{A,N-1})}
\epsilon_{  j_{A,1} j_{A,2}\cdots   j_{A,N-2} j_{A,N-1} i_A}.  \label{eq:CR4'}
\eeq
Note that $\mu_X^k$ is a $k$-th power of the matrix $\mu_X$, and in particular $(\mu_X^0)^{i}_{~j}=\delta^{i}_{~j}$.
There are also relations obtained by applying  the cyclic permutation $A \rightarrow B \rightarrow C \rightarrow A$ to the above relations.

Since there is one to one correspondence between the sets $(\tr \mu_X^2,\cdots,\tr \mu_X^N)$ and $(v_{X,2},\cdots,v_{X,N})$,
we also have $v_{A,k}=v_{B,k}=v_{C,k}$, or more concisely
\beq
P_A(x)=P_B(x)=P_C(x).\label{eq:CR1'}
\eeq
Because of this relation, we may just write $v_{X,k}$ and $P_X(x)$ as $v_{k}$ and $P(x)$ by dropping the subscripts $A,B$ or $C$.
This is already used in writing \eqref{eq:CR3}.

\subsection{Derivation of chiral ring relations}\label{eq:deriveCR}
Here we explain how the chiral ring relations written down in the previous section are derived or guessed.
Some of the results have been obtained in \cite{Gaiotto:2008nz,Benini:2009mz}.

\paragraph{The relations on $\tr \mu^k$:}
The relations \eqref{eq:CR1} are derived in the same way as the derivation of
$\tr \mu_A^2=\tr \mu_B^2=\tr \mu_C^2$ in \cite{Benini:2009mz}, which we now review.
We consider the $T_N$ theory coupled to the following quiver,
\beq
\SU(N-1)-\SU(N-2)-\cdots-\SU(2).
\eeq
A subgroup $\SU(N-1) \subset \SU(N)_C$ is gauged by the above $\SU(N-1)$,  there are bifundamental hypermultiplets
between $\SU(k)$ and $\SU(k-1)~(k=N-1,N-2,\cdots,2)$,
and there is one additional flavor for $\SU(2)$ to make it superconformal.
 This is dual~\cite{Gaiotto:2009we} to the following linear quiver;
\beq
\SU(N)_A -\SU(N)_2-\cdots-\SU(N)_{N-1}-\SU(N)_B, \label{eq:linearquiver}
\eeq
where $\SU(N)_A$ and $\SU(N)_B$ are flavor symmetries and $\SU(N)_k~k=2,\cdots,N-1$ are gauge symmetries.
We denote $\SU(N)_1=\SU(N)_A$ and $\SU(N)_N=\SU(N)_B$ for simplicity.
There are bifundamentals $(q_{k})^{i_k}_{~i_{k+1}}$ and $(\tilde{q}_{k})_{~i_k}^{i_{k+1}}$ between $\SU(N)_k$ and $\SU(N)_{k+1}$.
The identification of $\SU(N)_B$ with $\SU(N)_N$ is done as $g \in \SU(N)_N \mapsto {}^tg^{-1} \in \SU(N)_B$
so that the fundamental and anti-fundamental representations are exchanged, e.g., $(q_{N-1})^{i_{N-1}}_{~i_{N}}=(q_{N-1})^{i_{N-1}i_{B}}$.

The $T_N$ operators $\mu_A$ and $\mu_B$ are identified with the following operators in the dual quiver,
\beq
\mu_A=(q_1 \tilde{q}_1)_A,~~~~~\mu_B={}^t(\tilde{q}_{N-1} q_{N-1} )_B,
\eeq
where the subscript in a bilinear like $(q \tilde{q})_X$ means that we take the adjoint representation of $\SU(N)_X$ constructed from that bilinear,
e.g., $( \tilde{q}_kq_{k})_{k+1}=\tilde{q}_k q_k -\frac{1}{N}\tr \tilde{q}_k q_k$.
Taking the direction of $\SU(N)_C$ which is singlet under $\SU(N-1)$ as $i_C=1$,  we identify the operators $Q^{i_A i_B 1}$ and $Q_{i_A i_B 1}$ as
\beq
Q^{i_A i_B 1} \propto (q_1 q_2\cdots q_{N-1})^{i_A}_{~i_N},~~~~~Q_{i_A i_B 1} \propto (\tilde{q}_{N-1} \cdots \tilde{q}_{2}\tilde{q}_{1})^{i_N}_{~i_A},
\eeq
where we used the rules between the indices $i_{N}$ and $i_B$ discussed above.

The superpotential of the linear quiver \eqref{eq:linearquiver} is given as
\beq
\sum_{k=2}^{N-1}\tr [\Phi_k (-(\tilde{q}_{k-1} q_{k-1})_k+(q_k \tilde{q}_k)_k)],
\eeq
where the minus sign in the first term comes from the fact that $q_{k-1}$ is in the anti-fundamental
representation of $\SU(N)_k$ and hence it couples to $-{}^t \Phi_k$.
Then the equations of motion give $(\tilde{q}_{k-1} q_{k-1})_k=(q_k \tilde{q}_k)_k$.
By using this and cyclicity of trace, we get
\beq
\tr \mu_A^k&=\tr [(q_1 \tilde{q}_1)_1]^k=\tr [( \tilde{q}_1 q_1)_2 ]^k=\tr[(q_2 \tilde{q}_2)_2]^k=\cdots \nonumber \\
&=\tr[( \tilde{q}_{N-1} {q}_{N-1})_N]^k=\tr ({}^t\mu_B)^k.
\eeq
Taking the traceless part of matrices, e.g., $(q_1 \tilde{q}_1)_1=q_1 \tilde{q}_1-\frac{1}{N} \tr q_1 \tilde{q}_1$,
does not spoil the trace cyclicity used here.

\paragraph{The relations on $\mu Q$:}
Next, let us consider relations \eqref{eq:CR2}, and \eqref{eq:CR2'}.
 We use the same quiver dual. We start from the following relation:
\beq
&(q_1 \tilde{q}_1)_1 q_{1}q_2\cdots q_{N-1}=q_1 (\tilde{q}_1 q_{1})_2 q_2\cdots q_{N-1}=q_1 (q_{2}\tilde{q_2})_2 q_2\cdots q_{N-1} \nonumber \\
&=\cdots=q_1 q_2\cdots q_{N-1} (\tilde{q}_{N-1}q_{N-1})_N.
\eeq
This equation is equivalent to $(\mu_A)^{i_A}_{~j_A} Q^{j_A i_B 1}=(\mu_B)^{i_B}_{~j_B} Q^{i_A j_B 1}$.
This relation should extend to \eqref{eq:CR2} for general $i_C $. The relation \eqref{eq:CR2'} is obtained in the same way.

\paragraph{The relation on $QQ$ :}
The relation \eqref{eq:CR3} is determined by the following argument.\footnote{It was Davide Gaiotto who originally came up with the argument a few years ago. The authors thank him for kindly allowing them to reproduce it here. }
Let us define  $R^{i_Ai_B}_{~~~~j_A j_B}:=Q^{i_A i_B i_C}Q_{j_A j_B i_C}$.
Consider $N$ dimensional linear spaces $V_{A,B}$ on which $\SU(N)_{A,B}$ acts respectively.
$R$ is a linear operator on $V_A\otimes V_B$. The operators
$\mu_A\otimes 1$ and $1\otimes  \mu_B$ also act on $V_A\otimes V_B$; we abbreviate them as just $\mu_A$ and $ \mu_B$.

Now, using the relation \eqref{eq:CR2} and \eqref{eq:CR2'}, it is easy to show that $R$ commutes with $\mu_A$ and $\mu_B$. Therefore
$R$ should be a polynomial in $\mu_A$ and $\mu_B$.  The dimension of $Q$ says that this is a polynomial of degree $N-1$.
Again by \eqref{eq:CR2} and \eqref{eq:CR2'},  we get $(\mu_A-\mu_B)R=0$.
Such an $R$ can be constructed nicely using \eqref{eq:CR1'}.
From the property of the characteristic polynomial, we have $P(\mu_A)=P(\mu_B)=0$.
Thus we can suppose
\begin{equation}
R \propto \frac{P(\mu_A)-P(\mu_B)}{\mu_A-\mu_B}\label{Rmu},
\end{equation}
where the right hand side is understood that we first calculate the polynomial $[P(x)-P(y)]/(x-y)$ with a later substitution of $x$ by $\mu_A$
and $y$ by $\mu_B$; as $\mu_A$ and $\mu_B$ commute as an operator acting on $V_A \otimes V_B$
there is no problem in this procedure.
Writing \eqref{Rmu} explicitly, we get \eqref{eq:CR3}. The coefficient of \eqref{Rmu} is absorbed in the normalization of $Q^{i_A i_B i_C}$.

\paragraph{The relations relating $Q_{i_Ai_Bi_C}$ and $Q^{i_Ai_Bi_C}$:}
Finally we discuss the relations \eqref{eq:CR4} and \eqref{eq:CR4'}.
Let us start with the case of $T_2$ theory, that is, the theory of a free single trifundamental chiral field $Q^{i_A i_B i_C}$.
In this theory, we define $Q_{i_A i_B i_C}\equiv -\epsilon_{i_A j_A}\epsilon_{i_B j_B}\epsilon_{i_C j_C}Q^{j_A j_B j_C}$,
where the convention for the totally anti-symmetric tensor in this paper is $\epsilon^{12}=\epsilon_{12}=1$.
Then we have
\beq
Q^{i_A j_B j_C}\epsilon_{j_B i_B} \epsilon_{ j_C i_C}=Q_{j_A i_B i_C}\epsilon^{i_A j_A} ,~~~
Q_{i_A j_B j_C}\epsilon^{j_B i_B} \epsilon^{j_C i_C}=-Q^{j_A i_B i_C}\epsilon_{i_A j_A}. \label{eq:T2CR4}
\eeq
For the $T_3$ theory, all the chiral ring relations are worked out in \cite{Gaiotto:2008nz}.
There is a relation of the form\footnote{The relations in \cite{Gaiotto:2008nz} are written in the $\SU(2) \times \SU(6) \subset E_6$ covariant form.
We need to rewrite them in $\SU(3)\times \SU(3) \times \SU(3)$ covariant way to get \eqref{eq:T3CR4}. }
\beq
Q^{i_A j_B j_C} Q^{j_A k_B k_C} \epsilon_{i_B j_B k_B}\epsilon_{i_C j_C k_C} \propto Q_{k_A i_B i_C} (\mu_A)^{(i_A}_{~\ell_B} \epsilon^{j_A) \ell_A k_A},
\label{eq:T3CR4}
\eeq
and a similar relation with upper and lower indices exchanged. We need a generalization of \eqref{eq:T2CR4} and \eqref{eq:T3CR4}
to $N>3$.

On a generic point of the Higgs branch, the constraint \eqref{eq:CR1} is solved up to complexified $\SU(N)_A \times \SU(N)_B \times \SU(N)_C$
transformations by the vevs
\beq
\mu_A=\mu_B=\mu_C=\diag(m_1,m_2,\cdots,m_N), \label{eq:classicalmu}
\eeq
with $\sum_i m_i=0$. Then, the relations \eqref{eq:CR2} and \eqref{eq:CR2'} says that the only nonzero entries of $Q^{i_A i_B i_C}$ and $Q_{i_A i_B i_C}$
are for $i_A=i_B=i_C$, which we write as
\beq
Q^{iii}=q^i,~~~Q_{iii}=q_i~~~~~(i=1,\cdots,N),\label{eq:classicalQ}
\eeq
with other components zero. The relation \eqref{eq:CR3} fixes the product $q^iq_i$ for each $i$ (note that no summation is involved):
\beq
q^i q_i=\prod_{j \neq i}(m_i-m_j). \label{eq:qqproduct}
\eeq

In terms of these ``gauge-fixed'' variables, \eqref{eq:T2CR4} gives us $q^1=q_2$,
while \eqref{eq:T3CR4} gives $q^1 q^2 \propto q_3(m_1-m_2)$ and its cyclic permutations.
Using \eqref{eq:qqproduct}, they are written in more symmetric way as
$q^1 q^2=(m_2-m_1)$ and $q^1 q^2 q^3 \propto (m_1-m_2)(m_1-m_3)(m_2-m_3)$.
From these, it is easy to guess a general relation of the form
\beq
q^1q^2\cdots q^N \propto \prod_{1\leq i<j \leq N}(m_i-m_j). \label{eq:reduceoneofNq}
\eeq
This equation is consistent with the scaling dimensions of the operators and the  Weyl group action.
Dividing this by \eqref{eq:qqproduct}, we get
\beq
q^1q^2\cdots q^{N-1}  \propto q_{N}  \prod_{i=1}^{N-1} \prod_{j=i+1}^{N-1}(m_i-m_j), \label{eq:fixedCR4}
\eeq
and its cyclic permutation.
We also have the version with $q^i \leftrightarrow q_i$.

The relation \eqref{eq:reduceoneofNq} is crucial for reproducing the correct dimension of the Higgs branch of the $T_N$ theory.
The complex dimension of the Higgs branch is given by $(3N+2)(N-1)$, 
as can be shown by the S-duality discussed above or using the result of \cite{Benini:2010uu}. 
On the other hand, the $\mu_{A,B,C}$ with the constraint \eqref{eq:CR1}
contributes to the dimension as $3(N^2-1)-2(N-1)=(3N+1)(N-1)$.
The remaining $N-1$ directions of the Higgs branch are provided by the $q^i$'s with 
the constraint \eqref{eq:reduceoneofNq}.\footnote{We believe that other $Q_{(k)}$ for $k=2,\cdots,N-1$ are determined
by $q^i$'s and $m_i$'s in a similar way as the $q_i$'s are determined by them in \eqref{eq:qqproduct}.}

One can check explicitly that the relations \eqref{eq:CR4} and \eqref{eq:CR4'} reproduce \eqref{eq:fixedCR4} and
its cyclic permutation and the version with $q^i \leftrightarrow q_i$, up to coefficients.
The coefficient of \eqref{eq:CR4} is absorbed in the relative normalization of $Q^{i_A i_B i_C}$ and $Q_{i_A i_B i_C}$,
while the coefficient of \eqref{eq:CR4'} is determined by the consistency with \eqref{eq:CR4} and \eqref{eq:qqproduct}.

As one can see from the above discussion, the relations \eqref{eq:CR4} and \eqref{eq:CR4'} are more speculative than others.
However, they play an important role when we discuss reduction of the $T_N$ theory to a bifundametal hypermultiplet
in appendix~\ref{appsec:ConstraintsClosingPunc} (see \eqref{eq:QNcomp1} and \eqref{eq:QNcomp2} where they are used),
and that can be regarded as an overall consistency check of \eqref{eq:CR4} and \eqref{eq:CR4'}.

\subsection{Chiral rings from superconformal index}
\label{sec:chiral}
  In the subsequent sections we will study the $T_{N}$ theory coupled to an $\CN=1$ vector multiplet
  by gauging, say, $\SU(N)_{C}$ flavor symmetry.
  To do that, we need the spectrum of the Higgs branch operators
  which are invariant under the gauged $\SU(N)_{C}$ symmetry.
  Here we will see they are generated by $\mu_A$ and $\mu_B$ only,
  with the constraints $\tr\mu_A^k =\tr \mu_B^k$ via the study of the Hall-Littlewood index.
  We also analyze the $\SU(N)_{C}$ invariant operators in the case with a generic puncture.

  The Higgs branch of a class-$\mathcal{S}$ theory can be characterized by the Hall-Littlewood index
  \cite{Gadde:2011uv} defined to be as follows\footnote{We changed the notation from \cite{Gadde:2011uv} as
    $(R,r) \rightarrow (I_{3}, - \frac{1}{2}R_{\CN=2})$.}
    \begin{equation}
    \cI^{HL}={\Tr}_{HL}(-1)^F\tau^{2(E-I_{3})},
    \end{equation}
  where ${\Tr}_{HL}$ denotes the trace restricted to states with $\delta_{1\pm}=E\pm 2j_1-2I_{3}+\frac{1}{2}R_{\CN=2}=0$
  and $E$ is conformal dimension.
  Note that for the Higgs branch operators $2(E-I_{3}) = E$.
  We consider the SCFTs obtained by compactifying $N$ M5-branes on a sphere with three punctures.
  A puncture is specified by a partition $\Lambda = (n_{1}, \ldots, n_{\ell})$
  with $n_{1} \geq n_{2} \geq \ldots \geq n_{\ell}$ and $\sum n_i=N$.
  The flavor symmetry of the puncture is ${\rm S}[\prod_{i} \U(r_{i})]$
  where $r_{i} = n^{T}_{i} - n^{T}_{i+1}$ and $\Lambda^{T} = (n_{1}^{T},\ldots,n_{n_{1}}^{T})$
  is the dual partition to $\Lambda$.
  The Hall-Littlewood index of the theory associated with a sphere with three generic punctures is
  expressed in terms of the Hall-littlewood polynomial $P^\lambda_{HL}(\bfa|\tau)$ of partition
  $\lambda=(\lambda_1,\lambda_2,\cdots,\lambda_{N-1})$,
  which are orthogonal polynomials with respect to the pairing
    \begin{equation}
    \int [d\bfa] \Delta_{HL}(\bfa) P^\lambda_{HL}(\bfa|\tau)P^{\lambda'}_{HL}(\bfa^{-1}|\tau)
     =     \delta^{\lambda\lambda'},\quad
    \Delta_{HL}(\bfa)
     =     (1-\tau^2)^{N}\Delta_{Schur}(\bfa)\hat{\CK}(\bfa), \label{HLmeasure}
    \end{equation}
  where $\Delta_{Schur}(\bfa)$ is the standard Haar measure.
  The Hall-Littlewood index is given by
    \begin{equation}
    \cI^{HL}(\bfa_A,\bfa_B,\bfa_C)
    =\frac{\prod^N_{j=2}(1-\tau^{2j})}{(1-\tau^2)^{-N-2}}\prod_{X}\hat{\CK}_{\Lambda_X}(\bfa_X)
    \sum_{\lambda_1\geqslant\lambda_2\geqslant\cdots\geqslant\lambda_{N-1}}
     \frac{\prod_{X}P^{\lambda}_{HL}(\bfa_X(\Lambda_X)|\tau)}{P^{\lambda}_{HL}(\tau^{N-1},\ldots,\tau^{1-N}|\tau)},
    \end{equation}
  where $\bfa_X(\Lambda_X)$ are flavor fugacities of the puncture $\Lambda_X$ and
    \bea
    \hat{\CK}_{\Lambda}(\bfa)
     =     \prod^{n_{1}}_{i=1}\prod^{n_i^{T}}_{j,k=1}\frac{1}{1- \tau^{n_j+n_k-2i+2} a_j/a_{k}}.
    \eea
  The flavor fugacities satisfy $\prod_{j=1}^{\ell} a_{j}^{n_{j}} = 1$.
  In what follows, let us choose the punctures $A$ and $C$ are maximal with $\SU(N)$ flavor symmetries,
  and define the following index
  by integrating out the fugacities corresponding to the $\SU(N)_C$ symmetry
    \begin{equation}
    \label{eq:SingIndDef}
    \cI^{sing}(\bfa,\bfb)=\int [d\bfc]\Delta_{Schur}(\bfc)\cI^{HL}(\bfa,\bfb,\bfc),
    \end{equation}
  where we defined $\bfa_{A}=\bfa$, $\bfa_{B}=\bfb$ and $\bfa_{C}=\bfc$.

  Let us first consider the $T_{N}$ theory with three maximal punctures.
  The fugacities $\bfa(\Lambda_{max})$ are just fugacities of $\SU(N)$ flavor group $a_1,a_2,\cdots,a_N$
  with constraint $\prod^N_{i} a_i=1$, so as $\bfb$ and $\bfc$.
  The $\hat{\CK}$ factor in this case is
    \bea
    \hat{\CK}(\bfa)=\frac{1}{(1-\tau^2)^{N}}\prod^N_{i\neq j}\frac{1}{1-\tau^2a_i/a_j}.
    \eea
  Using \eqref{HLmeasure}
  the relevant part of the integral (\ref{eq:SingIndDef}) is
    \bea
    (1-\tau^2)^{-N}\int [d\bfc]\Delta_{HL}(\bfc)P^{\lambda_1,\ldots,\lambda_{N-1}}_{HL}(\bfc|\tau).
    \eea
  The orthogonality of HL polynomial implies that the integral is zero
  except $\lambda_1=\lambda_2=\cdots=\lambda_{N-1}=0$ in which case the HL polynomial is a constant.
  Therefore, the $\cI^{sing}$ index is given simply by
    \begin{equation}
    \cI^{sing}
     =     \frac{\prod^N_{j=2}(1-\tau^{2j})}{(1-\tau^2)^{-2}}
           \hat{\CK}(\bfa)
           \hat{\CK(\bfb)}.
    \end{equation}
  This can be written in terms of the plethystic exponential as
    \begin{equation}
    \label{eq:IsingFullPunc}
    \cI^{sing}(\bfa,\bfb)
     =     PE \left[ \tau^2(\chi_{adj}(\bfa)+\chi_{adj}(\bfb))-\sum_{j=2}^N \tau^{2j} \right],
    \end{equation}
  where we used $\hat{\CK}(\bfa)(1-\tau^2) = PE[\tau^2\chi_{adj}(\bfa)]$
  and $\prod^N_{j=2}(1-\tau^{2j})=PE[-\sum_{j=2}^N \tau^{2j}]$.
  The first two terms $\chi_{adj}(\bfa)+\chi_{adj}(\bfb)$ inside the plethystic exponential correspond to
  polynomials generated by $\mu_A$ and $\mu_B$.
  Then the subtraction terms correspond to the relations $\tr \mu_A^k = \tr \mu_B^k$, $k=2,\ldots, N$.

  Let us then consider other examples where the puncture $A$ is not maximal while $B$ and $C$ kept intact.
  When the puncture $A$ is minimal associated with the Young diagram $\Lambda = (N-1,1)$
  with $\U(1)_{A}$ flavor symmetry,
  the theory is just a hypermultplet in the bifundamental
  of $\SU(N)_B$ and $\SU(N)_C$ and charged under the $\U(1)_A$.
  The resulting index is
    \bea
    \cI^{sing}
     =     (1-\tau^{2N})PE\left[\tau^2\left(\sum_{i=1}^Nb_i\right)\left(\sum_{j=1}^N\frac{1}{b_j}\right)\right]
       PE\left[\tau^N(a^N+\frac{1}{a^N})\right].
    \eea
  The three factors represent, from right to left, the baryons $B$ and the antibaryons $\tilde B$,
  the mesons $M$, and one relation $\det M-B\tilde{B}=0$.

  A possible generalization is the puncture $A$ associated with the partition $(N-k,1,\cdots,1)$
  which corresponds to the flavor symmetry ${\rm S}[\U(1) \times \U(k)]$.
  The corresponding index is
    \begin{equation}
    \label{eq:U1UkPunc}
    \prod_{j=N-k+1}^N(1-\tau^{2j})PE\left[\tau^2\chi_{adj}(\bfb)\right]PE\left[\tau^2\right]
    PE\left[\tau^2\chi_{adj}(\bfa)\right] PE\left[\tau^{N-k+1} \sum_{i=1}^{k}(\frac{a}{a_i}+\frac{a_i}{a})\right],
    \end{equation}
  where $a$ and $a_{i}$ are the fugacities of $\U(1)$ and $\U(k)$ satisfying $a^{N-k} \prod_{i=1}^{k} a_{i}=1$.
  This represents the following spectrum of the $\SU(N)_{C}$ invariants
    \begin{equation}
    \begin{array}{|c||c|c|c|c|}
    \hline
     & \Delta & \U(1) & \SU(k) & \SU(N)_B \\\hline \hline
    \mu_B & 2 & 0 & \cdot & {\rm adj} \\
    \mu^{0} & 2 & 0 & \cdot & \cdot \\
    \mu_{\SU(k)} & 2 & 0 & {\rm adj} & \cdot \\
    B & N-k+1 & {N}/{k} & \Box & \cdot \\
    \tilde{B} & N-k+1 & -{N}/{k} & \overline{\Box} & \cdot \\
    \hline
    \end{array}
    \label{spectrum}
    \end{equation}
  and $k$ constraints among them.

\section{Deformed moduli space}
\label{sec:deformed}

\subsection{Statement of the result}\label{sec:deformedresult}
  In this section, we consider $\CN=1$ $\SU(N)$ supersymmetric gauge theory
  obtained by gauging an $\SU(N)$ flavor symmetry of the $T_{N}$ theory.
  Let us denote the gauged symmetry by $\SU(N)_{C}$ and the remaining flavor symmetry of the $T_{N}$ theory by
  $\SU(N)_{A} \times \SU(N)_{B}$.
  As is shown in section~\ref{sec:chiral}, the chiral operators which are singlets
  under the $\SU(N)_{C}$ symmetry of the $T_{N}$ theory are $\mu_{A}$ and $\mu_{B}$ which satisfy the relation
    \bea
    \tr \mu_{A}^{k}
     =     \tr \mu_{B}^{k},
    \eea
  where $k=2, \ldots, N$.
  This describes the original moduli space of vacua before coupling to the gauge fields.

  Our proposal is that in the IR the theory is confined as in the case of $\SU(N) $ supersymmetric QCD (SQCD)
  with $N_f=N$ flavors~\cite{Seiberg:1994bz}:
  the effective theory consists of the gauge singlet fields $\mu_{A}$ and $\mu_{B}$.
  However the classical moduli space is deformed to the quantum one
    \bea
    \tr \mu^{k}_{A} - \tr \mu^{k}_{B}
     = N    \Lambda^{2N} \delta^{k N},
           \label{constraint}
    \eea
    where $\Lambda^{2N}$ is the holomorphic dynamical scale of $\SU(N)_C$. They can be concisely summarized as
    \beq
    P_A(x)=P_B(x)-\Lambda^{2N}, \label{eq:conscisedeformed}
    \eeq
    where $P_{A,B}(x)$ are the characteristic polynomials defined in \eqref{eq:charateristicP}.

  The global non-anomalous symmetry of the UV theory is $\SU(N)_{A} \times \SU(N)_{B} \times \U(1)_{R}$
  where $\U(1)_R$ is the same as the $\U(1)_{R (\CN=2)}$ symmetry of the $\CN=2$ $R$-symmetry of the $T_{N}$ theory.
  Another $R$-symmetry of the $T_N$ theory, $\U(1)_{2I_3}$ is anomalous
  and the dynamical scale $\Lambda^{2N}$ has charge $2N$ under it.
  The chiral operators $\mu_{A, B}$ have charges $0$ and $2$
  under the $\U(1)_{R}$ and $\U(1)_{2I_{3}}$ symmetries respectively.
  Also the gaugino in the $\CN=1$ vector multiplet has charge $1$ under each $\U(1)_{R}$ and $\U(1)_{2I_3}$.

  Then, up to coefficient, \eqref{constraint} is the only possibility consistent with these symmetries.
  We assume that the coefficient of $\Lambda^{2N}$ is nonzero and absorb it in the definition of $\Lambda^{2N}$.
  Note that we cannot have a dynamical superpotential of $\mu_A$ and $\mu_B$ since they are neutral under
  the $R$-symmetry $\U(1)_R$.
  Therefore the SUSY cannot be broken spontaneously.
  These discussions are completely parallel to the SQCD case \cite{Seiberg:1994bz}.
  The fact that the coefficient of $\Lambda^{2N}$ in  \eqref{constraint} is nonzero will be supported by the fact that we can reproduce the result of
  SQCD from reduction of the $T_N$ theory as discussed in section~\ref{subsec:TNtoSQCD}, if and only if the coefficient is nonzero.

  On a generic point of the quantum moduli space, the flavor symmetry is broken to $\U(1)^{2(N-1)} \times \U(1)_{R}$,
  where the $\U(1)^{2(N-1)}$ come from the Cartan subalgebra of $\SU(N)_{A,B}$.

  \subsection{Higgsing from $\SU(N)$ to $\SU(N-1)$}
  It is possible to reduce this theory to the $\SU(N-1)$ theory coupled to the $T_{N-1}$ theory,
  as the $\SU(N)$ SQCD with $N_f=N$ flavors of quarks
 can be reduce to the $\SU(N-1)$ SQCD with $N-1$ flavors by giving vevs to quarks.
 We can consider a large vev
 \beq
 \vev{\mu_A}=\vev{\mu_B}=\vev{\mu_C}=m \diag (1,\cdots,1,-N+1),~~~\vev{Q^{NNN}Q_{NNN}}=(-Nm)^{N-1}, \label{eq:TN-1}
 \eeq
 where we used \eqref{eq:qqproduct}. By this vev, the gauge symmetry is broken to $\SU(N-1)$, which is coupled
 to the $T_{N-1}$ theory.\footnote{The fact that the $T_{N-1}$ theory is obtained in this way from the $T_N$ theory may be seen as follows from the 
 M-theory point of view.
 The $T_{N}$ theory is realized as $N$ coincident M5 branes wrapped on a Riemann sphere with three punctures.
 Then, the vev \eqref{eq:TN-1} for $\mu$'s is interpreted as seperating one M5 brane from the other $N-1$ M5 branes in a certain direction.
 These $N-1$ M5 branes give the $T_{N-1}$ theory.
 } 
 We also have free Nambu-Goldstone multiplets $(\mu_X)^{i_X}_{~N}$ and $(\mu_X)^{N}_{~i_X}~(X=A,B)$
 for $i_X \leq N-1$ and a modulus field $m$ which are decoupled from the strong dynamics.

 The deformed moduli constraint for this $\SU(N-1)$ theory can be obtained in the following way.
 In \eqref{eq:conscisedeformed} we redefine $x'=x-m$, and take the limit $m \to \infty$ while $x'$ and $\Lambda^{2N}/m$ fixed.
 Then we get
 \beq
 P'_A(x')=P'_B(x')-\frac{\Lambda^{2N}}{Nm},
 \eeq
 where $P'_X$ is the characteristic polynomial for $(\mu'_X)^{i_X}_{~j_X} \equiv (\mu_X-\vev{\mu_X})^{i_X}_{~j_X}$, where $1\leq i_X, j_X \leq N-1$.
 This is clearly the deformed moduli constraint for the $\SU(N-1)$ theory with the low energy dynamical scale
 given as $\Lambda^{2N}/Nm$.

\subsection{Check of the anomaly}
  As a check of the deformation of the moduli space \eqref{constraint}, we check 't Hooft anomaly matching here.
  To compare the anomaly coefficients of the UV and IR theories,
  we choose the vevs as $\vev{\tr \mu_{A}^{N}} = N\Lambda^{2N}$ and $\vev{\mu_{B}} = 0$
  where the $\SU(N)_{B}$ flavor symmetry is unbroken, while the $\SU(N)_{A}$ symmetry is broken to $\U(1)^{N-1}$.
  The number of the massless Nambu-Goldstone fields from $\mu_{A}$ is $N^{2}-N$.

  Let us first compare the anomaly coefficient $\tr T^{a}_{B} T^{b}_{B} R$,
  where $T_{B}^{a}$ and $R$ are the generators of the $\SU(N)_{B}$ and $\U(1)_R$.
  In the UV theory this anomaly is calculated from the flavor central charge of the $T_{N}$ theory
    \bea
    \tr T^{a}_{B} T^{b}_{B} R
     =   - N \delta^{ab}.
    \eea
  This agrees with the anomaly of the IR theory $- t({\rm adj}) \delta^{ab}$
  to which only the $\mu_{B}$ field contributes,
  where we used the normalization of the Dynkin index $t$ such that $t(\square) = \frac{1}{2}$ and $t({\rm adj}) = N$.

  Next let us consider the $\tr R^{3}$ anomaly.
  In the UV theory the contribution of the $T_{N}$ theory can be computed by using
  $\tr R_{\CN=2}^{3} = 2(n_{v}-n_{h}) = - 3N^{2} + N + 2$ \cite{Gaiotto:2009gz, Chacaltana:2012zy}.
  Adding the contribution of the gaugino, we get
    \bea
    \tr R^{3}
     =   - 2N^{2} + N + 1.
    \eea
  This agrees with that of the IR theory.
  Note that the reduction of the components of $\mu_{A}$, due to the quantum constraints,
  is crucial in the matching.
  Also the anomaly $\tr R$ is the same as the above because $\tr R_{\CN=2} = \tr R_{\CN=2}^{3}$,
  and the $R$ charges of the gaugino and the fermionic partner of the low energy fields are $1$ and $-1$.

  Finally, we consider the anomaly coefficients involving unbroken $\U(1)^{N-1}$ symmetries from $\SU(N)_{A}$.
  Let $(U_{i})^k_{~\ell}$ ($i=1,\ldots,N-1$, and $k,\ell=1,\ldots,N$) be generators of these $\U(1)$'s
  such that the only non-zero components are $(U_{i})^i_{~i} = - (U_{i})^{i+1}_{~i+1} = 1$.
  In the IR theory, the components $(\mu_{A})^{i}_{~i+1}$ and $(\mu_{A})^{i+1}_{~~i}$ have
  charge $\pm 2$, and the components $(\mu_{A})^{i}_{~k}$, $(\mu_{A})^{k}_{~i}$, $(\mu_{A})^{i+1}_{~~k}$
  and $(\mu_{A})^{k}_{~i+1}$ ($k \neq i,i+1$)
  have charge $\pm 1$. Then the anomaly under $U_i^2R$ in IR is given as
  \beq
  -2\cdot (\pm 2)^2-4(N-2) \cdot (\pm 1)^2=-4N
  \eeq
  In the UV theory, this can be computed as
    \bea
    \tr U_{i}^{2} R
     =   - 4N,
    \eea
  where the factor $4$ comes from the normalization of $U_i$, $\tr U_i^2=2$,
  compared with $\tr T^a T^b=\frac{1}{2}\delta^{ab}$.
  Therefore the UV and IR anomalies match.
  The anomalies $\tr \U(1)_{i} R^{2}$ and $\tr \U(1)_{i}^{3}$ vanish.
  The only other nonzero anomalies are $\tr U_i U_{i+1}R$,
  which can be computed similarly with the result $\tr U_i U_{i+1}R=2N$.

\subsection{Higgsing to SQCD and others}
\label{subsec:TNtoSQCD}
  It is known \cite{Benini:2009gi,Chacaltana:2012zy} (see also \cite{Gaiotto:2008ak} in a different context)
  that a puncture specified by a partition described in \cite{Gaiotto:2009we} can be obtained
  from a maximal puncture by giving a nilpotent vev to the corresponding moment map (say $\mu_A$).
  Let us consider an embedding $\rho :\SU(2) \to \SU(N)$ given
  by $\Box \to {\bf n_1}+{\bf n_2}+\cdots+{\bf n_\ell}$,
  where ${\bf n_k}~(k=1,\cdots,\ell)$ are $n_k$ dimensional (spin $(n_k-1)/2$) representations of $\SU(2)$.
  This embedding $\rho$ is specified by the partition $\Lambda=(n_{1},\ldots,n_{\ell})$
  with $n_{1} \geq \ldots \geq n_{\ell}$, {\it i.e.}, $\rho$ has a Jordan block structure
  such that $k$-th block has the size $n_k\times n_k$.
  Then, by giving a vev $\vev{\mu_A}=\rho_{\Lambda}(\sigma^+)$,
  we get a puncture specified by the partition and some Nambu-Goldstone multiplets
  associated with the symmetry breaking of $\SU(N)_A$ by the vev $\rho(\sigma^+)$.
  Under such embedding the adjoint representation of $\SU(N)$ decomposes as
    \begin{equation}
    \label{eq:GenPuncDecomp}
    \mathrm{adj}
     =     \left[\bigoplus_{\alpha=1}^{\ell} \bigoplus_{j=1}^{n_{\alpha} -1} V_{j} \right]
           \oplus (\ell-1) V_{0}\oplus
           2 \left[\bigoplus_{\alpha < \beta} \bigoplus_{k=1}^{n_{\beta}}
           V_{ \frac{n_{\beta} + n_{\alpha} - 2k}{2} }\right]
           \equiv\bigoplus_j R_{j} \otimes V_{j},
    \end{equation}
  where $V_j$ is the spin $j$ representation of $\SU(2)$ and $R_j$ is the flavor symmetry representation.
  The first two terms and the last term come from the diagonal and off-diagonal blocks, respectively.

  In particular, the $T_N$ theory is reduced to a bifundamental hypermultiplet
  ${\cal Q}^{i_B i_C}, \tilde{{\cal Q}}_{i_B i_C}$ together with Nambu-Goldstone multiplets
  by closing one puncture into the minimal ($\U(1)$) puncture with a vev of $\mu_A$ as
    \beq
    \vev{\mu_A}=\rho_{\star}( \sigma^+),
    \label{eq:nilpotentvevMain}
    \eeq
  where $\rho_\star$ is the embedding $\SU(2) \to \SU(N)$ given by $\Box \to {\bf (N-1)}+{\bf 1}$.
  In this subsection we sketch the reduction of the deformed moduli constraints proposed
  in section \ref{sec:deformedresult} to that of $\SU(N)$ SQCD with $N_f=N$ flavors by the above process.
  The detailed derivation is left to appendix \ref{appsec:ConstraintsClosingPunc}.

  By the vev \eqref{eq:nilpotentvevMain}, the $\SU(N)_A$ flavor symmetry is broken to $\U(1)$ generated
  by $T_{\U(1)}={\rm diag}(1,\cdots,1,-(N-1))$. Then $\mu_A$ is decomposed as
    \beq
    \mu_A
     =     \rho_{\star}( \sigma^+)+\sum_{j=1}^{N-2} \sum_{m=-j}^{j}T_{j,m,0} \mu_A^{j,m,0}+T_{\U(1)} \mu_A^{0,0,0}
         + \sum_{m=-\frac{N-2}{2}}^{\frac{N-2}{2}} T_{\frac{N-2}{2},m,\pm N} \mu_A^{\frac{N-2}{2},m,\pm N},
    \eeq
  where $T_{j,m,q}$, $T_{\U(1)}$ and $\rho_{\star}$ satisfy the following commutation relation
    \beq
    [\rho_\star (\sigma^3), T_{j,m,q}]
     =     m T_{j,m,q}, ~~
    [\rho_\star (\sigma^\pm), T_{j,m,q}]
     \propto  T_{j,m \pm 1,q}, ~~
    [T_{\U(1)},T_{j,m,q}]
     =     q T_{j,m,q}.
           \label{comm}
    \eeq
  One can factor out the Nambu-Goldstone modes $\mu_A^{j,m,q}$ with $m>-j$
  corresponding to the broken $\SU(N)_A$ by reparametrization $\mu_A = G_{A} \mu'_A G_A^{-1}$,
  where $G_{A}$ is an element of the complexified $\SU(N)_A$ group containing the Nambu-Goldstone multiplets.
  All the holomorphic constraint equations are covariant under the complexified $\SU(N)_A$,
  and hence $G_A$ drops out from them.
  That is, the Nambu-Goldstone multiplets do not enter into the constraints.
  With these multiplets out, $\mu_A$ can be parametrized into the following matrix form
    \beq
    \mu_A
     \sim  \left( \begin{array}{cccc|c}
           \mu_A^{0,0,0} & (\rho_\star(\sigma^+))^1_{~2}  & 0 & 0&0 \\ 
           \vdots & \ddots &  \ddots & 0 & \vdots \\ 
           \mu_A^{N-3,-(N-3),0} &    & \ddots & (\rho_\star(\sigma^+))^{N-2}_{~N-1}  & 0 \\
           \mu_A^{N-2,-(N-2),0} &\mu_A^{N-3,-(N-3),0}  & \ldots& \mu_A^{0,0,0} & \mu_A^{\frac{N-2}{2},-\frac{N-2}{2}, N}
           \\\hline
           \mu_A^{\frac{N-2}{2},-\frac{N-2}{2},- N} & 0 & \ldots & 0 & -(N-1)\mu_A^{0,0,0}
           \end{array}
           \right),\label{eq:matrixformAMain}
    \eeq
  where we have neglected order one coefficients of components containing $\mu_A^{j,-j,0}~(j \geq 1)$.

  Now we impose the constraints $\tr \mu_A^k-\tr \mu_B^k = N\Lambda^{2N} \delta^{kN}$.
  Solving the constraints with $k=2,\ldots,N-1$ determines $\mu_A^{k-1,-(k-1),0}$
  in terms of $\mu_A^{0,0,0}$ and $\mu_B$.
  The constraint with $k=N$ gives the relation
    \beq
    \det (-(N-1)\mu_A^{0,0,0} {\bf 1}-\mu_B)
    + C(\rho) \mu_A^{\frac{N-2}{2},-\frac{N-2}{2}, N}\mu_A^{\frac{N-2}{2},- \frac{N-2}{2}, -N}
     =     \Lambda^{2N},
           \label{eq:redconstraintMain}
    \eeq
  where $C(\rho) = \prod_{k=1}^{N-2} (\rho_\star(\sigma^+))^k_{~k+1}$.

  In SQCD, we have mesons $\CM^{i_B}_{~j_B}={\cal Q}^{i_B i_C}\tilde{{\cal Q}}_{j_B i_C}$
  and baryons $\CB_+=\det {\cal Q}$ and $\CB_-=\det \tilde{\cal Q}$.
  The deformed moduli constraint is given by $\det \CM-\CB_+ \CB_-=\Lambda^{2N}$.
  Eq.~(\ref{eq:redconstraintMain}) is the same as this SQCD constraint by the following identification:
    \beq
    \hat{\CM} \equiv \CM-\frac{{\bf 1}}{N} \tr \CM
     =     \mu_B, ~~
    \tr \CM
     =     N(N-1) \mu_A^{0,0,0}, ~~
    \CB_\pm
     =     c_\pm \mu_A^{\frac{N-2}{2},-\frac{N-2}{2}, \pm N},
           \label{eq:idTNSQCDMain}
    \eeq
  where $c_\pm$ are constants satisfying $c_+ c_- =(-1)^{N-1}C(\rho)$.
  This can be checked explicitly as demonstrated in appendix \ref{appsec:ConstraintsClosingPunc}.
  Here let us shortly see this is reasonable one.
  Before gauging $\SU(N)_C$, the theory has ${\cal N}=2$ supersymmetry.
  The operators $\hat{\CM}$ and $\tr \CM$ are the holomorphic moment maps
  of the $\SU(N)_B$ and $\U(1)$ symmetries respectively,
  so the first two identifications of (\ref{eq:idTNSQCDMain}) are expected from the mapping of these moment maps.
  Note that the operator $\tr \CM$ maps to $\tr (\mu_A T_{\U(1)})=N(N-1)\mu_A^{0,0,0}$,
  which explains the coefficient $N(N-1)$ in the second equation of (\ref{eq:idTNSQCDMain}).
  Up to coefficients, the identification of the baryon operators is also easy to expect just from the $\U(1)$ charges.
  
  The reproduction of the SQCD result is highly nontrivial.
    Recall that the usual SQCD has $\SU(N) \times \SU(N) \times \U(1)$ flavor symmetry
    which almost fixes the form of the constraint.
    However, only the subgroup $\SU(N)_B \times \U(1)$ of the SQCD symmetry group is manifest in the reduction of the $T_N$ theory
    to the SQCD, and other symmetries should be realized as accidental symmetries at low energies.
    The manifest symmetry $\SU(N)_B \times \U(1)$ is not enough at all to determine the form of the constraint \eqref{eq:redconstraintMain}
    which is consistent with the SQCD.

  It is also interesting to consider the closure of one puncture of the $T_N$ theory to other generic one.
  This is an $\SU(N)$ gauge theory coupled to the SCFT associated with a sphere with two maximal punctures
  and one generic puncture.
  Here we consider the case with the next-to-maximal puncture $\Lambda = (N-k,1,\cdots,1)$
  with the flavor symmetry $\U(1) \times \SU(k)$
  and derive the constraints of the quantum moduli space of this theory.
  As in the previous section, the index indicates that the spectrum is given by \eqref{spectrum} with $k$ relations.
  $\mu_A$ can be parametrized (after factoring out Nambu-Goldstone modes),
  in terms of the gauge invariants in \eqref{spectrum},
  in a similar block diagonal form
    \begin{equation}
    \mu_A\sim
    \left(
      \begin{array}{ccccc|c}
      \mu^{0} & \rho^1_2 & 0 & 0 & 0 & 0 \\
      \mu^1 & \mu^{0} & \rho^2_3 & 0 & 0 & \vdots  \\
      \vdots & \ddots & \ddots &\ddots & \vdots & \vdots \\
      \mu^{N-k-2} & & \ddots & \mu^{0} & \rho^{N-k-1}_{N-k} & 0  \\
      \mu^{N-k-1} & \mu^{N-k-2} & \cdots & \mu^1 & \mu^{0} & B  \\
      \hline
      \tilde{B} & 0 & 0 & \cdots & 0 & -\frac{N-k}{k}\mu^{0}I_{k\times k}+\mu_{\SU(k)}
      \end{array}
    \right),
  \end{equation}
  where $B = ( B_1, B_2, \cdots, B_k )$,
  ${}^{t}\tilde{B} = ( \tilde{B}^{1}, \tilde{B}^{2}, \cdots, \tilde{B}^{k} )$,
  and we suppressed unnecessary indices.
  A similar argument as the previous case gives the constraints in matrix form.
  The $\SU(k)$ flavor symmetry guarantees that there are only $k$ independent constraints.
  Here we present the expression of $k=2$ case:
  the constraint $P_B(\mu_A) =  \Lambda^{2N} I_{N \times N}$ gives
  \begin{equation}
   P_B\left(-\frac{N-2}{2}\mu^0I_{2\times2}+ \mu_{\SU(2)} \right)+C(\rho)\left(\mu_{\SU(2)} \tilde{B} B+  \tilde{B} B \mu_{\SU(2)} \right)
     =   \Lambda^{2N} I_{2\times2},
  \end{equation}
  where $C(\rho)=\prod^{N-3}_{k=1}\rho^k_{k+1}$.

  With a similar procedure, for example, we should be able to derive the deformed moduli space of $\cN=1$ $\SU(N)$ theory
  coupled to two fundamental flavors together with one flavor of two-index anti-symmetric tensor \cite{Nanopoulos:2009xe}.

\section{Adding one flavor}
\label{sec:Nc+1}

\subsection{Statement of the result}
\label{subsec:Nc+1}
  We add one massless flavor $q$ and $\tilde{q}$ to the setup of the previous section.
  So we have one $T_N$ theory, an $\SU(N)$ vector multiplet gauging one $\SU(N)$ flavor symmetry of the $T_N$ theory, and this additional flavor $q$ and $\tilde q$.

  The theory has the global symmetry
  $\SU(N)_{A} \times \SU(N)_{B} \times \U(1)_{B} \times \U(1)_F\times \U(1)_{R}$.
  Here, the two $\SU(N)$'s come from the flavor symmetry of the $T_{N}$ theory,
  and the $\U(1)_{B}$ acts only on $q$ and $\tilde{q}$ with charge $1$ and $-1$.
  The $\U(1)_F$ is given by
    \begin{equation}
    F
     =     R_{\cN=2}-2I_3+NA,
    \end{equation}
  where $A$ is the generator of the $\U(1)$ symmetry under which both $q$ and $\tilde q$ have charge $1$.
  Finally, the $R$-symmetry is given by
    \bea
    R
     =     R_{\CN=2} + A.
    \eea
  All these are anomaly free. Charges of various fields are summarized in the following table:
    \begin{equation}
    \begin{array}{|c||c|c|c|}
    \hline
              & \U(1)_{B} & \U(1)_{F} & \U(1)_{R} \\
    \hline \hline
    {q}         & 1 & N       & 1 \\
    \tilde{ q} & -1& N       & 1 \\
    \mu_{X}   & 0 & -2      & 0 \\
    \hline
    \end{array}
    \label{elementarycharges}
    \end{equation}

  The one-loop beta function is the same as that of the $\CN=1$ SQCD with $N_{f} = N + 1$.
  Thus we expect that the low energy theory exhibits the confinement
  and is described by the gauge invariant operators with dynamically generated superpotential as in \cite{Seiberg:1994bz}.
  Let us first list the gauge invariant operators of this theory:
    \beq
    &{B}_{i_A i_B}
     =     Q_{i_A i_B i_C}q^{i_C}, ~~~
    \tilde{B}^{i_A i_B}
     =     Q^{i_A i_B i_C} \tilde{q}_{i_C}, \nonumber \\
    &\mu_X~~(X=A,B), ~~~
    {M}_k
     =     \tilde{q} \mu_C^k q~~(k=0,1,\cdots,N-2).
    \eeq
  The $\U(1)$ charges of these gauge invariants are summarized as
    \begin{equation}
    \begin{array}{|c||c|c|c|}
    \hline
              & \U(1)_{B} & \U(1)_{F} & \U(1)_{R} \\
    \hline \hline
    {B}         & 1 & 1       & 1 \\
    \tilde{ B} & -1& 1       & 1 \\
    \mu_{X}   & 0 & -2      & 0 \\
    {M}_{k}     & 0 & -2k+2N  & 2 \\
    \hline
    \end{array}
    \label{charges}
    \end{equation}
  Note that the meson ${ M}_{N-1}=\tilde{q} \mu_C^{N-1} q$ is excluded. This is because we can express this operator using
  other gauge invariant operators due to the relation \eqref{eq:CR3}.

  From the global symmetry, the low energy superpotential is given as
    \beq
    W
     =    - \frac{1}{\Lambda^{2N-1}}\left[\sum_{k=0}^{N-2} {M}_k f_{N-k}(\mu_A,\mu_B) -\tr [({B}\tilde{B})_A\mu_A]
         + \tr [(\tilde{B} { B})_B \mu_B] \right],
           \label{superpotentialNc+1}
    \eeq
  where $f_{N-k}(\mu_A,\mu_B)$ is an $\SU(N)_A \times \SU(N)_B$ invariant polynomial of degree $N-k$
  given below, and $\Lambda$ is the dynamical scale. The coefficient of the term $\tr [({B}\tilde{ B})_A\mu_A]$ can be absorbed
  in the definition of $\Lambda^{2N-1}$, while that of $ \tr [(\tilde{ B} { B})_B \mu_B] $ is determined by later consistency.

  The $f_{N-k}$ is determined to be
    \beq
    f_{N-k}(\mu_A,\mu_B)
     =     \sum_{\ell=0}^{N-k-2} \frac{ v_{\ell}(\mu_A,\mu_B)}{N-k-\ell} (\tr \mu_A^{N-k-\ell} - \tr \mu_B^{N-k-\ell} ),
           \label{ef}
    \eeq
  where $v_\ell(\mu_A,\mu_B)$ is a function of $\mu_A$ and $\mu_B$ of degree $\ell$
  which coincides with $v_{A,\ell}$ and $v_{B,\ell}$ when $\tr \mu^k_A=\tr \mu^k_B~(k=2,\cdots,N)$.
  This $f_{N-k}$ is determined such that the equations of motion give the ``classical'' constraints
  which are the chiral ring relations of the $T_{N}$ theory.
  Indeed, the equations of motion with respect to ${M}_{k}$ give the relation $\tr \mu_A^k=\tr \mu_B^k$ ($k=2,3,\cdots,N)$, which is \eqref{eq:CR1}.
  The equation of motion of $\tilde{B}$ gives
  \beq
  0=(\mu_A)^{j_A}_{~i_A} { B}_{j_A i_B}-(\mu_B)^{j_B}_{~i_B} { B}_{i_A j_B}. \label{eq:Beqmotion}
  \eeq
  This is derived from \eqref{eq:CR2'} in the UV.
  The equations of motion with respect to $\mu_A$ is
    \beq
    0
     =     \sum_{k=0}^{N-2} \sum_{\ell=0}^{k}
           v_{\ell} {M}_{k-\ell} \left(\mu_A^{N-k-1} - \frac{\bf{1}}{N} \tr \mu_A^{N-k-1} \right)^{i_A}_{~j_A}
         - { B}_{j_A i_B}\tilde{ B}^{i_A i_B}
         + \frac{\delta^{i_A}_{~j_A}}{N} { B}_{k_A k_B}\tilde{B}^{k_A k_B},
           \label{relationNc+1}
    \eeq
  where we have used $\tr \mu_A^k=\tr \mu_B^k$.
  This can be obtained from the relation \eqref{eq:CR3}.

  The consistency with the theory in the previous section is seen as follows:
  By adding a mass term of $q$ and $\tilde{q}$, the IR superpotential is just \eqref{superpotentialNc+1}
  plus $m { M}_0$, which can be seen by using the global symmetries.
  Thus we can easily get
  $\tr \mu_A^{N}-\tr \mu_B^{N} =N m \Lambda^{2N-1}$
  and $\tr \mu_A^{k}-\tr \mu_B^{k}=0~(k=2,\cdots, N-1)$.
  These are the quantum constraints of the previous case.

 In the above analysis, we have not completely determined the function $v_\ell (\mu_A, \mu_B)$.
 However, the ambiguity in this function can always be absorbed by redefining ${M}_k$ as
    \beq
    { M}_k
     \to
    {M}'_k={ M}_k+\sum_{\ell=1}^{k} h_{\ell}(\mu_A,\mu_B){ M}_{k-\ell} ,\label{4.10}
    \eeq
  where $h_{\ell}(\mu_A,\mu_B)$ is an arbitrary function which vanishes under the condition $\tr \mu_A^{k}=\tr \mu_B^{k}$.
  This new ${ M}'_k$ is equivalent to the original ${ M}_k$ as an element of chiral rings.\footnote{
  If we add a tree level superpotential like $\lambda M_k$, we get $\tr \mu_A^{N-k}-\tr \mu_B^{N-k} \propto \lambda \Lambda^{2N-1}$,
  and hence $M'_k$ is different from $M_k$ by an amount proportional to $\Lambda^{2N-1}$ via \eqref{4.10}.
  However, one should note that there is a general ambiguity in
  the definition of composite operators like $M_k=\tilde{q} \mu_C^k q$. We need some regularization to define composite operators.
  (If we could consider them just as classical objects, we could have proved e.g. $\det \CM-\CB_+ \CB_-=0$ in SQCD by
  classical algebraic manipulation, which is not the case.)
  In the regularization, there is an ambiguity in the choice of (finite part of) counterterms.
  We require that these ambiguous terms are: (1) consistent with global symmetries and holomorphy,
  (2) going to zero when $\Lambda^{2N-1} \to 0$, since there is a canonical way of defining composite operators of chiral fields
  when we turn off
  gauge interactions (by using point splitting regularization).
  In these criteria, $M'_k$ is as good as $M_k$ as a definition of
  the ambiguous operator $\tilde{q} \mu_C^k q$. There is no physical principle to select one of them.
  }
 Therefore, the remaining ambiguity is not important as long as we do not try to determine the effective K\"ahler potential.

\subsection{Check of the anomaly}
  At the origin of the moduli space, the full global symmetry is unbroken.
  We compare the anomaly coefficients at this point.
  In the UV theory, these are computed by using the anomalies of the $T_{N}$ theory
  as in the previous section.
  Here we only list the results
    \bea
    & &
    \tr T^{a}_{X} T^{b}_{X} R
     =     \tr T^{a}_{X} T^{b}_{X} F
     =   - N \delta^{ab}, ~~~
    \tr B^{2} R=0,~~~
    \tr B^{2} F
     =     2N^2,
    \nonumber \\
    & &
    \tr R
     =     \tr R^{3}
     =     - 2 N^{2} + N + 1, ~~~
    \tr F
     =     - N^2 + N+ 2,
    \nonumber \\
    & &
    \tr F^{3}
     =     2 N^4 + 4 N^3 - 12 N^2 + 8,
    \nonumber \\
    & &
    \tr R^{2} F
     =    - 3 N^2 + N + 2,~~~
    \tr R F^{2}
     =     4N^3/3 - 6 N^2 + 2N/3 + 4
    \eea
  where $X = A, B$.
  Note that we have used the anomaly of the $T_{N}$ theory: $\tr R_{\CN=2} (2I_{3})^{2} = 4 N^3/3 - 3 N^2 - N/3 + 2$ and
  $\tr R_{\CN=2}^{3} =\tr R_{\CN=2} = - 3N^{2} + N + 2$
  for the calculation of $\tr F^{3}$, $\tr F R^{2}$ and $\tr F^{2} R$.
  One can check that these agree with the anomalies of the IR theory
  computed from \eqref{charges}.

\subsection{Higgsing to SQCD}
\label{sec:}
  In this subsection we will see that the superpotential \eqref{superpotentialNc+1} is reduced,
  by closing the puncture as in section~\ref{subsec:TNtoSQCD}, to the effective superpotential of SQCD with $N_{f} = N+1$,
    \bea
    \Lambda^{2N-1} W_{{\rm SQCD}}
     =     \CB \CM \tilde{\CB} - \det \CM,
    \eea
  where the baryon and meson operators are written
  in terms of quark fields $\CQ^{I i_{C}}$ and $\tilde{\CQ}_{I i_{C}}$ ($I = 1,\ldots,N+1$) as
    \beq
    \CB_{I}
    & =   \frac{1}{N!}  \CQ^{I_{1} i_{C,1}} \cdots \CQ^{I_{N} i_{C,N}} \epsilon_{I_{1} \ldots I_{N} I}
           \epsilon_{i_{C,1} \ldots i_{C,N}},~~~ \nonumber \\
    \tilde{\CB}^{J}
    & =   \frac{1}{N!}  \tilde{\CQ}_{J_{1} i_{C,1}} \cdots \tilde{\CQ}_{J_{N} i_{C,N}} \epsilon^{J_{1} \ldots J_{N} J}
           \epsilon^{i_{C,1} \ldots i_{C,N}},~~~~~
    \CM^{I}_{~J}
     =     \CQ^{I  i_C} \tilde{\CQ}_{J i_C}.
           \label{baryon}
    \eeq
  Here ${\cal Q}^{I=N+1}=q$, $\tilde{\cal Q}_{I=N+1}=\tilde{q}$, and ${\cal Q}^I,~\tilde{Q}_I$ for $I=i_B=1,\cdots,N$
  are as in \eqref{eq:bifundamentalID}.

  Let us consider the first term of the superpotential \eqref{superpotentialNc+1}.
  For $k=1,\ldots,N-2$, by using \eqref{eq:muApower}
  we can see that $M_{k}$ and $\mu_A^{N-k-1,-(N-k-1),0}$ form a mass term in the superpotential
  and they can be integrated out.
  The equations of motion of these massive fields give $\tr \mu_A^{N-k}=\tr \mu_B^{N-k}$ for $k=1,\cdots,N-2$.
  Then, the remaining term in the first term of \eqref{superpotentialNc+1} is
    \bea
    M_{0} f_{N}
     =     M_{0} \sum_{l=0}^{N-2} \frac{v_{l}}{N-l} (\tr \mu_{A}^{N-l} - \tr \mu_{B}^{N-l}).
    \eea
  Again the terms with $l\neq 0$ vanish by using the constraints.
  Under the condition $\tr \mu_A^{k}=\tr \mu_B^{k}$ for $k=2,\cdots,N-1$,
  we get $\frac{1}{N} (\tr \mu_{A}^{N} - \tr \mu_{B}^{N}) = -P_A(x)+P_B(x)$, where $x$ is arbitrary.
  Taking $x=-(N-1)\mu_A^{0,0,0}$ as in \eqref{eq:specialxvalue}, we obtain
    \beq
    M_0 f_N
    &=(-1)^{N}\CM^{N+1}_{~ N+1}\left(\det_{(N \times N)} \CM -\CB_{N+1} \tilde{\CB}^{N+1} \right),
    \label{eq:superpiece1}
    \eeq
  where we have used identification \eqref{eq:idTNSQCDMain} and $\CM^{N+1}_{~ N+1}=M_0$.

  Next, let us consider the other two terms of \eqref{superpotentialNc+1}.
  Because of the vev of $\mu_{A}$, the fields $B_{i_A i_B}~(i_A=1,\cdots,N-2)$
  and $\tilde{B}^{i_A i_B}~(i_A=2,\cdots,N-1)$ become massive.
  Therefore we need to integrate them out.
  The equations of motion with respect to these massive fields are
    \beq
    &\tilde{B}^{i_{A} j_{B}} (\mu_{B})^{i_{B}}_{~j_{B}}
     =     \tilde{B}^{j_{A} i_{B}} (\mu_{A})^{i_{A}}_{~j_{A}}~~~(i_A=1,\cdots,N-2), \label{eq:massiveeq1}\\
    &B_{i_{A} i_{B}} (\mu_{B})^{i_{B}}_{~j_{B}}
     =     B_{j_{A} i_{B}} (\mu_{A})^{j_{A}}_{~i_{A}}~~~(i_A=2,\cdots,N-1). \label{eq:massiveeq2}
    \eeq
  By using these and after some algebra which is detailed in appendix \ref{appsec:ConstraintsClosingPunc},
  we can rewrite the last two terms of \eqref{superpotentialNc+1} as
    \bea
    (-1)^{N-1} \Big[\CB_{ i_B} \tilde{ \CB}^{ i_B} \CM^{i_B}_{~j_B}
    + \CB_{ i_B} \CM^{ i_B}_{~N+1} \tilde{\CB}^{N+1}
    +  \CM^{N+1}_{~ i_B} \tilde{\CB}^{i_B} \CB_{N+1}  \nonumber \\
    + \CM^{N+1}_{~i_B} \det\CM (\CM^{-1})^{i_B}_{~j_B}\CM^{j_B}_{~N+1} \Big],
      \label{eq:superpiece2}
    \eea
    where the identification of operators is as in \eqref{eq:Nf=N+1ID}.
  By combining \eqref{eq:superpiece1} and \eqref{eq:superpiece2},
  the superpotential \eqref{superpotentialNc+1} is thus reduced to
    \beq
    \frac{(-1)^{N-1}}{\Lambda^{2N-1}}\left[   \det_{(N+1) \times (N+1)} \CM   - \CB_I \CM^I_{~J} \tilde{\CB}^J \right],
    \eeq
which is the SQCD superpotential.

\section{$\cN=1$ Coulomb branch}
\label{sec:Coulomb}

  So far, we have considered the $T_{N}$ theory where the $\SU(N)_{C}$ flavor symmetry is gauged.
  In this section we will see that further gauging of additional $\SU(N)$ flavor symmetries
  leads to abelian Coulomb phase, whose low energy holomorphic coupling matrix is encoded
  in  an $\CN=1$ Seiberg-Witten curve~\cite{Intriligator:1994sm}.
  After analyzing the case of the $T_{N}$ theory coupled to $\CN=1$ $\SU(N)^{3}$ vector multiplets
  and with or without flavors, we generalize this to $\CN=1$ quiver gauge theory.
  Our analysis is a generalization of the previous studies
  \cite{Intriligator:1994sm,Kapustin:1996nb,Kitao:1996np,Giveon:1997gr,Csaki:1997zg,Gremm:1997sz,Giveon:1997sn,
  Tachikawa:2011ea} of conventional theories in the Coulomb phase to non-conventional ones.

\subsection{$T_N$ theory coupled to  $\SU(N)^3$}
  Let us consider the $T_{N}$ theory coupled to $\cN=1$ $\SU(N)^3$ vector multiplets.
  Recall that in the theory considered in section \ref{sec:deformed},
  the independent operators are $\mu_{A}$ and $\mu_{B}$.
  Since the $\SU(N)_{A}$ and $\SU(N)_{B}$ symmetries are gauged, the gauge invariant operators are
  the $v_{k}$
  ($k=2,\ldots,N$).
  There are another Higgs branch operators $Q_{i_{A}i_{B}i_{C}}$ and $Q^{i_{A}i_{B}i_{C}}$ of the $T_{N}$ theory.
  As discussed in section~\ref{eq:deriveCR}, a generic point of the moduli space at the ``classical'' level is given by
  the diagonal vevs \eqref{eq:classicalmu} and \eqref{eq:classicalQ}.
  Therefore, the gauge symmetry is broken to $\U(1)^{2(N-1)}$, represented in terms of the gauge fields $A_{X}^{\alpha}$
  of $\U(1)^{\alpha}_{X} \subset \SU(N)_{X}$ ($\alpha = 1, \ldots, N-1$) by
    \bea
    A_{A}^{\alpha} + A_{B}^{\alpha} + A_{C}^{\alpha}
     =     0.
    \eea
  Thus the theory is in the Coulomb phase.
  Classically there are codimension one singular loci on the moduli space parameterized by $v_{k}$
  where some of the W-bosons become massless and the gauge symmetry is enhanced.

  To consider quantum theory, let us turn on the dynamical scales $\Lambda_{A,B,C}$.
  In the regime $\Lambda_{C} \gg \Lambda_{A,B}$, we can first use the result of section \ref{sec:deformed}
  and think of the system as $\SU(N)_A$ coupled to $\mu_A$, together with $\SU(N)_B$ coupled to $\mu_B$,
  with the constraints \eqref{constraint}.
  Thus, each ingredient is the $\CN=2$ $\SU(N)$ pure SYM theory.
   It is clear that there are $2(N-1)$ abelian vector multiplets on a generic point on the moduli space.
  The classical singularities are split into singularities where massless charged particles appear
  as in \cite{Seiberg:1994rs,Klemm:1994qs,Argyres:1994xh}.
  The constraint $\tr \mu^{N}_{A} - \tr \mu_{B}^{N} = N\Lambda^{2N}_{C}$ says that
  the singular loci for $\SU(N)_A$ and $\SU(N)_B$ are generically separated at the quantum level.

 Naively, in the limit $\Lambda_{C} \gg \Lambda_{A,B}$, we just have two separate $\SU(N)$ gauge groups with an adjoint field for each gauge group.
 However, the deformed moduli space of section~\ref{sec:deformed} has a Wess-Zumino-Witten term.\footnote{This is because
 we can reduce this theory to
 the $\SU(2)$ theory coupled to $T_2$ (i.e., a trifundamental chiral multiplet) by the process described in section~\ref{sec:deformed},
 and this $\SU(2)$ theory contains the Wess-Zumino-Witten term~\cite{Manohar:1998iy}. }
 As discussed in detail in \cite{Tachikawa:2011ea} for the case of $\SU(2)$, this
 makes the dynamics much more complicated and interesting.

 To avoid the complication due to the Wess-Zumino-Witten term, we follow the procedure in \cite{Tachikawa:2011ea}.
  We first add one flavor to the $\SU(N)_C$ gauge group with mass $m$,
  and consider the regime $\Lambda_C \gg \Lambda_{A,B}$.
  At the energy scale around $\Lambda_{C}$ the $\SU(N)_{C}$ theory is confined with the superpotential
  as in section \ref{sec:Nc+1}.
  This theory does not have a Wess-Zumino-Witten term since the topology of the moduli space is trivial (in the limit $m \to 0$).
  We can regard $\mu_A$ and $\mu_B$ as adjoint chiral fields of $\SU(N)_A$ and $\SU(N)_B$, respectively,
  and ${B}, \tilde{ B}$ as bifundamental fields of $\SU(N)_A \times \SU(N)_B$.
  Furthermore,
  the superpotential \eqref{superpotentialNc+1} is similar to the one in an ${\cal N}=2$ theory
  except for the first term.
  The curve of this $\CN=2$ theory is written as \cite{Witten:1997sc}
    \beq
    \Lambda^{2N}_A t^3-P_A(w) t^2+P_B(w) t -\Lambda^{2N}_B
     =     0,
    \eeq
  where $P_{A,B}(w)=\det (w\cdot {\bf 1} - \mu_{A,B})$.
  The equations of motion with respect to $M_k$ give the relation $P_A=P_B- \Lambda^{2N}_C$,
  where the low energy dynamical scale $\Lambda^{2N}_C$ is defined as $m\Lambda^{2N-1}_C$.
  However, this relation is valid only in the limit $\Lambda_C \gg \Lambda_{A,B}$.
  We propose that the exact relation between $P_A$ and $P_B$ is given as
    \beq
    P_A
     =    P+\Lambda^{2N}_A,~~~~~
    P_B
     =    P+\Lambda^{2N}_C+\Lambda^{2N}_B,
          \label{eq:assumption1}
    \eeq
  where $P=w^N+\sum_{k=2}^{N} v_k w^{N-k}$ is a polynomial of $w$ of degree $N$.
  Then, the curve is written as
    \beq
    F(w,t) \equiv w^{N}
     +     \sum_{k=2}^{N} V_{k}(t) w^{N-k}=0,
    \eeq
  where $V_{k} (t) = v_{k}$ for $k=2,\ldots,N-1$ and
    \beq
    V_{N}(t)
     =   v_{N} - t \Lambda^{2N}_A-\frac{\Lambda^{2N}_B}{t} - \frac{\Lambda^{2N}_C}{t-1} .
           \label{firstcurve}
    \eeq
  The exact form (\ref{eq:assumption1}) has been obtained so that
  $V_{N}$ becomes a meromorphic function on a Riemann sphere with simple poles at $t=0,1$ and $\infty$.
  This should be the case because there is an obvious symmetry under the exchange of $A$,$B$ and $C$.
  The $\CN=1$ curve is the $N$-sheeted cover of the base sphere with three punctures.

  As mentioned above, the holomorphic gauge coupling matrix is identified with the period matrix of this curve.
  Let us see the genus of the curve. There are two types of branch points on the base Riemann sphere.
  One is at the poles of $V_{N}(t)$. The other is at the points
  where $F(w,t)=0$ has double roots, that is, $F(w,t) = 0$ and $\frac{\partial F(w,t)}{\partial w} = 0$.
  We can find $3(N-1)$ points of this type.
  The branching number of each simple pole is $N-1$, i.e.~$N$ sheets meet at these points.
  Meanwhile that of each double root point is $1$.
  By using the Riemann-Hurwitz relation, the genus of the curve $g'$ is
    \bea
    g'
     =     N(g - 1) + 1 + \frac{B}{2}
     =     2(N-1),
    \eea
  where $g$ is the genus of the base curve, which is zero in this case,
  and $B$ is the total branching number. This coincides with the number of massless $\U(1)$ fields.

  Let us study what happens when we decouple  some of the gauge groups.
  By decoupling the $\SU(N)_{A}$ vector multiplet by setting $\Lambda_{A} = 0$,
  the theory is $\SU(N)_{B} \times \SU(N)_{C}$ theory coupled to the $T_{N}$ theory.
  The theory is still in the Coulomb phase with $\U(1)^{N-1}$ gauge groups.
  The curve is simply \eqref{firstcurve} with $\Lambda_{A}=0$, and this is the same curve
  as the pure ${\cal N}=2$ SYM theory.
  Now, further decoupling the $\SU(N)_{B}$ vector multiplet by $\Lambda_{B} = 0$,
  the theory goes back to the one considered in section \ref{sec:deformed}.
  Indeed, the curve degenerates to a genus zero curve, indicating that all massless photons decouple.

\subsection{$T_N$ theory coupled to $\SU(N)^3$ with a number of additional flavors}
  We can further generalize the above theory by introducing some more fundamental flavors of quarks.
  Let us add $N_X(<N)$ flavors of quarks $q^X_I, \tilde{q}^X_I~(I=1,\cdots, N_X)$
  which are in the (anti)-fundamental representation of $\SU(N)_X$, where $X=A,B$.
   The superpotential for these flavors is taken as
    \beq
    W
     =     \sum_{X=A,B}\sum_{I=1}^{N_X}
           \left( m^X_I  \tilde{q}^X_I q^X_I + \lambda^X_I  \tilde{q}^X_I \mu_X  q^X_I  \right),
           \label{afterintegrate}
    \eeq
   We will soon give a motivation for this form of the superpotential.
   Again, the theory is similar to one studied in \cite{Witten:1997sc} in the limit $\Lambda_C \gg \Lambda_{A,B}$.
   The curve is
    \beq
    \Lambda^{2N-N_A}_A Q_A(w) t^3-P_A(w) t^2+P_B(w) t -\Lambda^{2N-N_B}_B Q_B(w) \label{eq:wittentype2}
     =     0,
    \eeq
  where $Q_{X}(w)=\prod_{I=1}^{N_{X}} (\lambda_I^{X}w+m_I^{X})$.
  The dependence of the curve on $m_I^{A,B}$ and $\lambda_I^{A,B}$ is determined
  by using spurious flavor symmetries
  such as a $\U(1)$ spurious symmetry
  \beq
  q^{X}_I \to \e^{i \theta_{I,X}} q^{X}_I , ~~ m^{X}_I \to e^{-i \theta_{I,X}} m^{X}_I ,~~
  \lambda^{X}_I \to e^{-i \theta_{I,X}} \lambda^{X}_I, ~~ \Lambda^{2N-N_X}_X \to e^{i \theta_{I,X}} \Lambda^{2N-N_X}_X  .
  \eeq
  We neglect possible order one coefficients.
  As in the previous case, we assume the relation between $P_A$ and $P_B$ is given as
  $P_A=P+\Lambda^{2N-N_A}_A Q_A$, $P_B=P+\Lambda^{2N}_C+\Lambda^{2N-N_B}_B Q_B$.
  The curve is then
    \beq
    P(w)
     =     t \Lambda^{2N-N_A}_A Q_A(w) + \frac{\Lambda^{2N-N_B}_B Q_B(w)}{t}+\frac{\Lambda^{2N}_C}{t-1}.
    \eeq
  This is again written as the $N$-sheeted cover of the sphere with three punctures
    \bea
    w^{N} + \sum_{k} V_{k}(t) w^{N-k}=0, \label{eq:TNcurvewflavor}
    \eea
  but in the present case $V_{k}(t)$ are as follows:
  at $t=0$, $V_{k}$ with $k \geq N-N_{B}$ has a simple pole,
  at $t=1$, only $V_{N}$ has a simple pole
  and at $t = \infty$, $V_{k}$ with $k \geq N- N_{A}$ has a simple pole.

  Notice that when $\lambda^{X}_I =0$,
  we recover the curve without flavors by identifying
  $\Lambda_{X}^{2N}=\Lambda^{2N-N_{X}}_{X} \prod_{I=1}^{N_{X}} m_I^{X}$.
  This means that in the theory without the cubic term in \eqref{afterintegrate}
  the addition of the flavors does not change the form of the $\CN=1$ curve.
  Note also that in the limit where all the flavors are massless the $V_{k}$ functions
  behave such that at $t=\infty$ only $V_{N-N_{A}}$ has a simple pole
  and at $t=0$ only $V_{N-N_{B}}$ has a simple pole.

  So far we have considered the theories as purely $\CN=1$.
  However, we can obtain them from ${\cal N}=2$ theories by mass deformations.
  Let us take the gauge multiplets as ${\cal N}=2$ vector multiplets and introduce the superpotential,
      \bea
    W
     =   \sum_{X=A,B,C} \left( M_X \tr \Phi_X^2  + \tr \Phi_X \mu_X \right) +\sum_{X=A,B}  \sum_{I=1}^{N_X} \left(m^X_I  \tilde{q}^X_I q^X_I
         +   \tilde{q}^X_I  \Phi_{X} q^{X}_{I}  \right),
    \eea
    where $\Phi_X$ are the adjoint chiral multiplets.
   The ${\cal N}=2$ supersymmetry is broken only by the adjoint mass terms.
  Assuming that the mass parameters $M_{X}$ are large enough such that the classical analysis is valid,
  we integrate $\Phi_{X}$ out and get terms like $\tr \mu_{X}^{2}$, $ (\tilde{q}^X_I q^{X}_{I})^{2}$
  and $ \tilde{q}^X_I \mu_{X}  q^{X}_{I} $ with coupling constant $1/M_{X}$.
  Since there is the relation $\tr \mu_{A}^2=\tr \mu_{B}^2=\tr \mu_{C}^2$, the first kind of terms is cancelled
  by setting $\sum_{X} 1/M_{X} = 0$, which is necessary for the moduli space not to be lifted.
  The second kind of terms can be shown to be irrelevant for the curve \eqref{eq:wittentype2} by spurious symmetry argument.
  The third kind of terms is precisely the ones in \eqref{afterintegrate} with $\lambda^X_I \to 1/M_X$.

\subsection{Generalization}

  We consider the following class of $\CN=1$ gauge theories in the Coulomb phase as a generalization of the previous model.
  Instead of a single copy of $T_N$ theory, we use a class of theories~\cite{Gaiotto:2009we} specified by a Riemann surface with several maximal punctures,
  and we gauge the flavor symmetries associated with the punctures by ${\cal N}=1$ vector multiplets.

\begin{figure}
  \begin{center}
    \includegraphics[scale=0.27]{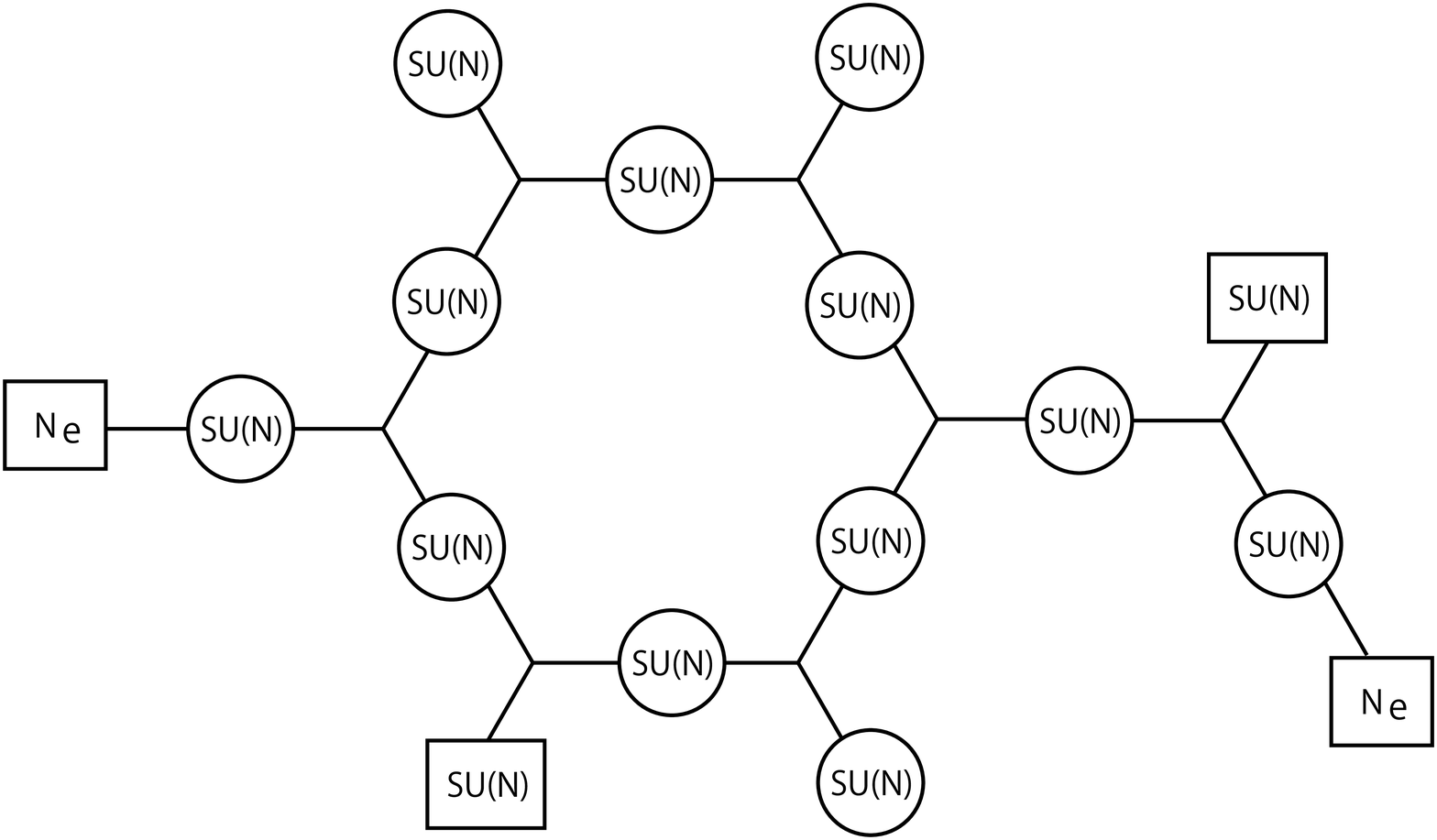}
    \caption{An example of generalized models.}\label{fig:quiver}
  \end{center}
\end{figure}

  The class is specified by a graph consisting of trivalent vertices connecting circles or boxes as in figure~\ref{fig:quiver}.
  The $\CN=1$ theory can be read off from a given graph as follows:
    \begin{itemize}
    \item Label vertices by $v$ and circles by $i$ or $e$.
    \item To each trivalent vertex $v$, we introduce a copy of $T_{N}$ theory denoted as $T_N^{(v)}$.
          If it is connected to circles $i,j$ and $k$, it has operators $\mu_{v,i}$, $\mu_{v,j}$ and $\mu_{v,k}$, etc.
    \item To each circle $i$ connected to two vertices $v$ and $v'$,
          we associate an $\CN=2$ $\SU(N)_i$ vector multiplet.
          The superpotential is taken as
            \bea
            W
             =     \tr (\mu_{v,i} - {}^t\mu_{v',i}) \Phi_{i}.
            \eea
          Note that the second term is $-{}^t \mu_{v',i}$ instead of just $\mu_{v',i}$.
          This comes from our choice of the embedding
          $g \in \SU(N)_i \mapsto  (g,~{}^tg^{-1})  \in \SU(N)_{v,i} \times \SU(N)_{v',i}$.
    \item To each box with $\SU(N)$ written inside it, we associate a flavor $\SU(N)$ symmetry.
    \item To each box with $N_{e}$ inside it, we introduce $N_{e}$ flavors of quarks ${q}_{e,I}$
          and $\tilde{q}_{e,I}$ with mass $m_{e,I}$ $(I=1,\cdots,N_e)$.
          We only consider the case $N_e <N$.
    \item To each circle $e$ connected to only one vertex $v$ and possibly to a box with $N_{e}$,
          we introduce a mass-deformed $\CN=2$ $\SU(N)_e$ vector multiplet.
          The superpotential is
          \beq
          W= M_{e} \tr \Phi_{e}^2 + \tr \mu_{v,e} \Phi_{e}+\sum_{I=1}^{N_{e}}\left[ \tilde{q}_{e,I} \Phi_{e} q_{e,I}+m_{e,I}\tilde{q}_{e,I}  q_{e,I} \right].
         \eeq
         We require the mass parameters $M_e$ satisfy
         \beq
         \sum_e \frac{1}{M_e}=0.\label{eq:masstune}
         \eeq
    \end{itemize}

Suppose that there are $g$ loops in the graph and $n$ $\SU(N)_e$ gauge fields which are coupled to only one trivalent vertex.
Our proposal is that the ${\cal N}=1$ curve of this class of models which encode the holomorphic coupling matrix of low energy $\U(1)$ fields
is given by
\beq
w^N+\sum_k V_k(t)w^{N-k}=0,\label{eq:generalcurve}
\eeq
where $V_k$ are meromorphic functions on a genus $g$ Riemann surface with $n$ punctures $\CC_{g,n}$.
At each puncture, $V_k$ with $k \geq N-N_e$ have simple poles, and the residues of these poles
are given in terms of dynamical scales of $\SU(N)_e$, $\Lambda_e^{2N-N_e}$, and the masses of quarks,
$m_{e,I}$.

This is a generalization of the result in \cite{Tachikawa:2011ea} for $\SU(2)$ with trifundamentals
to $\SU(N)$ with copies of $T_N$ theory and some flavors of quarks.\footnote{
Aside from the introduction of flavors of quarks, there is a few minor changes
from \cite{Tachikawa:2011ea}. First, the adjoint fields $\Phi_i$ connecting two
vertices was given mass terms in \cite{Tachikawa:2011ea}. In the case $g \geq 1$, there is actually a flat direction of $\Phi_i$.
However, we can show that the vevs of $\Phi_i$ do not enter in the curve because
they are charged under the $R_{{\cal N}=2}$ symmetry while the curve is neutral under this symmetry.
Another difference is that the massive adjoint was not introduced for gauge groups $\SU(N)_e$ connected to one vertex.
If there are no flavors of quarks as in \cite{Tachikawa:2011ea}, this difference does not matter as long as \eqref{eq:masstune} is satisfied.
However, massive adjoint fields are important to get the couplings $\tilde{q}_{e,I} \mu_{v,e} q_{e,I}$ as explained in the previous subsection.}
Note the similarity to the family of curves Gaiotto has written down for ${\cal N}=2$ theories~\cite{Gaiotto:2009we}
\beq
x^N+\sum_k \phi_k x^{N-k}=0,
\eeq
where $\phi_k$ are $k$-th differentials on a Riemann surface.
In our case, they are replaced by functions $V_k$ on a Riemann surface.

Now we perform checks of our proposal. Some of the checks of \cite{Tachikawa:2011ea}
are extended straightforwardly to the present case, so we only study a few basic properties here.

\paragraph{Number of $\U(1)$ gauge groups and moduli}
First let us study the moduli space and massless $\U(1)$ fields at the `classical' level, i.e.~before discussing the quantum effects from the gauge fields.
If a vertex $v$ is connected to circles labeled as e.g., $i,j$ and $e$,  we have the $T_N^{(v)}$ constraints
\beq
\tr \mu_{v,i}^k=\tr \mu_{v,j}^k=\tr \mu_{v,e}^k.
\eeq
We also have the equations of motion of $\Phi_i$, $\mu_{v,i} = {}^t\mu_{v',i}$.
Up to gauge and flavor transformations, these equations are solved by giving the same diagonal vevs to all the $\mu$ operators as
\beq
\mu_{v,i}=\mu_{v,e}=\diag(m_1,\cdots,m_N),
\eeq
where $\sum_k m_k=0$.
Also, as discussed in section~\ref{eq:deriveCR}, the operators $Q_{(v)}$ should also have diagonal vevs as in \eqref{eq:classicalQ}.
Then the massless gauge fields are given by
$A_{i}^{\alpha}$ of $\U(1)^{\alpha}_{i} \subset \SU(N)_{i}$ ($\alpha = 1, \ldots, N-1$)
with the constraints at each vertex $v$,
\beq
(-1)^{h(v,i)}A_{i}^{\alpha}+(-1)^{h(v,j)}A_{j}^{\alpha}+(-1)^{h(v,e)}A_{e}^{\alpha}=0, \label{eq:photonconstraint}
\eeq
where $h(v,i)=0$ or $1$ if $\SU(N)_i$ is embedded in $\SU(N)_{v,i}$ as $g \mapsto g$ or $g \mapsto {}^t g^{-1}$, respectively.
This has the form of current conservation in an electric circuit at each vertex. Therefore, when $n \geq 1$,
the number of massless photons are given as
\beq
N_{\U(1)}=(N-1)(n+g-1).\label{eq:numberofphotons}
\eeq
The products $(Q_{(v)})^{kkk} (Q_{(v)})_{kkk}~(k=1,\cdots,N)$ and $\prod_{k=1}^{N}(Q_{(v)})^{kkk} $ are determined by $m_k$
as in \eqref{eq:qqproduct} and \eqref{eq:reduceoneofNq}, while other $N-1$ unconstrained fields $(Q_{(v)})^{kkk} $ are absorbed by
the linear combination of gauge fields appearing in \eqref{eq:photonconstraint} to form massive vector fields.
Therefore the moduli space is $N-1$ dimensional which is spanned by $m_k$ or equivalently $v_k~(k=2,\cdots,N)$.

Let us compare the above field theory results to the properties of the curve.
When  all the residues of poles of $V_k$ are fixed, the only remaining freedom is to change the constant parts of $V_k \sim v_k$.
These $N-1$ constants represent the moduli fields, consistent with the field theory analysis.

The number of photons should be compared with the genus of the curve.
The genus can be calculated as follows. The equation \eqref{eq:generalcurve} has $N$ solutions as a polynomial equation of $w$,
which we denote as $w_k(t)~(k=1,\cdots,N)$.
At the puncture corresponding to $\SU(N)_e$,
$V_k$ with $k \geq N-N_e$ have simple poles, and hence the behavior of $w_k$ near this pole is given as
\beq
w_k(t) \sim
\left\{
\begin{array}{ll}
 {\rm finite} & (1 \leq k \leq N_e)\\
 (t-t_e)^{-1/(N-N_e)}& (N_e+1 \leq k \leq N)
\end{array}
\right. , \label{eq:singularw}
\eeq
where $t_e$ is the position of the puncture. Therefore $N-N_e$ sheets meet at this point.
Next, let us determine the number of points at which \eqref{eq:generalcurve} has double roots.
Such points are the zeros of the discriminant,
\beq
D(t)=\prod_{k < \ell} (w_k(t)-w_\ell(t))^2.
\eeq
This discriminant $D(t)$ is a meromorphic function on the Riemann surface.
At the singular point, the behavior of $D(t)$ is determined by using \eqref{eq:singularw} as $D(t) \sim t^{-(N+N_e-1)}$.
Since the total of degrees of poles of a meromorphic function is the same as the total of degrees of zeros,
$D(t)$ has generically $\sum_e (N+N_e-1)$ zero points. Therefore, the genus of the curve $g'$ is determined by the Riemann-Hurwitz relation as
\beq
g'  =N(g-1)+1+\frac{1}{2} \sum_e \left[ (N-N_e-1)+(N+N_e-1) \right]
 =N_{\U(1)}+g.
\eeq
This is larger than \eqref{eq:numberofphotons} by $g$. This excess was already present in
e.g., \cite{Witten:1997sc,Gaiotto:2009we,Tachikawa:2011ea}.
The cycles pulled back from the base Riemann surface has a trivial monodromy when we change the moduli $v_k$.
They decouple from the rest of the cycles, and the remaining $g'-g=N_{\U(1)}$ cycles give us nontrivial
monodromy and the coupling matrix of massless photons in the field theory.

\begin{figure}
  \begin{center}
    \includegraphics[scale=0.28]{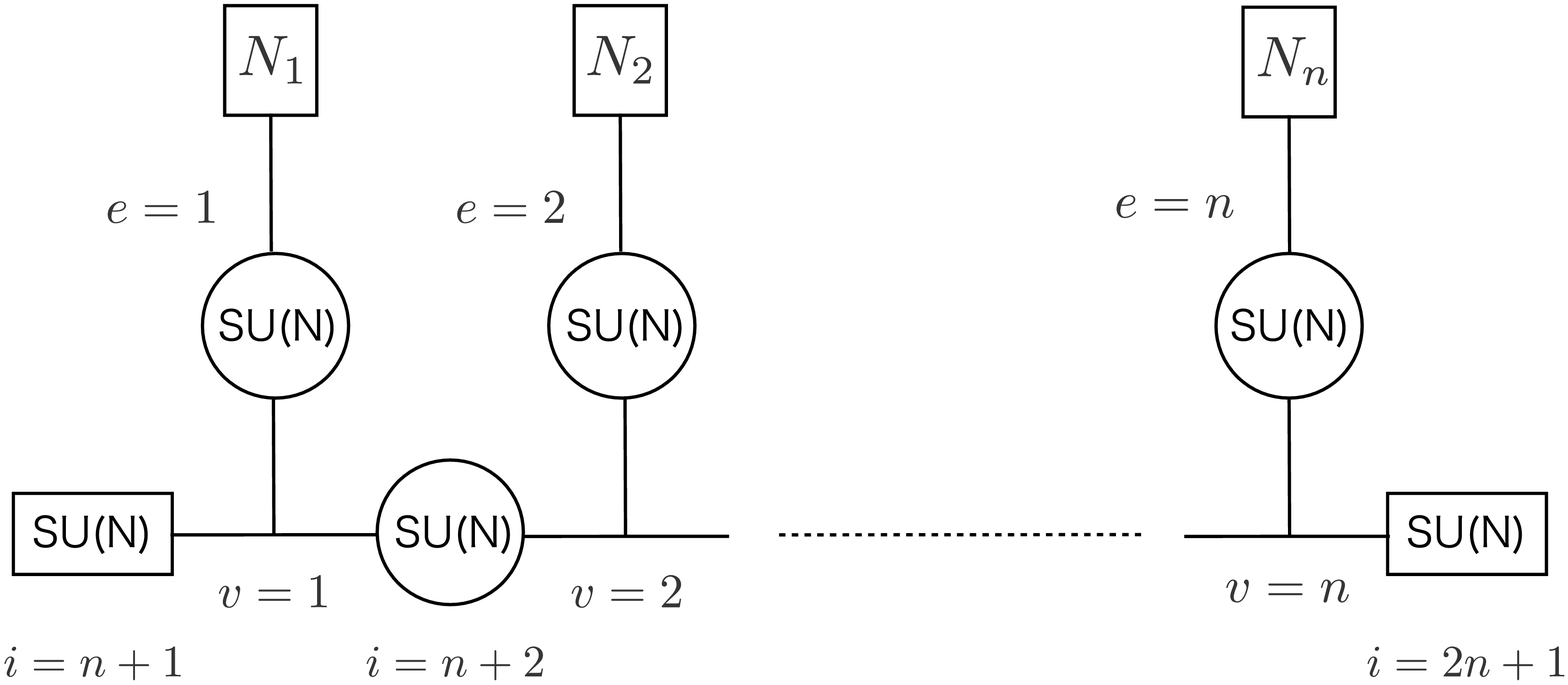}
    \caption{A genus-zero graph. A genus-one graph can be obtained by gauging the diagonal
    subgroup of $\SU(N)_{n+1}$ and $\SU(N)_{2n+1}$.}\label{fig:genusone}
  \end{center}
  \end{figure}

\paragraph{Conditions on dynamical scales}
  Let us consider the linear quiver theory as in figure \ref{fig:genusone}.
  Here we focus on the case with $N_{e} = 0$ or $1$.
  We consider the limit where all the dynamical scales $\Lambda_{e}^{2N}$ are large
  such that the $\SU(N)_{e}$ gauge theories ($e=1, \ldots, n$) confine.
  When $N_{e} = 0$ the low energy theory is described by
    \bea
    \tr \mu_{e,n+e}^{k} - \tr \mu_{e,n+e+1}^{k}
     =     N \Lambda_{e}^{2N} \delta^{kN}.
    \eea
  When $N_{e} = 1$ the relevant superpotential is given by
    \bea
    W
    &=&   - \frac{1}{\Lambda_{e}^{2N-1}} \left( \sum_{k} M_{e,k} f_{e,N-k} - \tr B_{e}\tilde{B}_{e} \mu_{e,n+e}
         + \tr \tilde{B}_{e} B_{e} \mu_{e,n+e+1} \right) + m_{e} M_{e,0} + \frac{1}{M_{e}} M_{e,1}.
           \nonumber \\ &&
    \eea
  where $M_{e,k}$, $B_{e}$ and $\tilde{B}_{e}$ are the $\SU(N)_{e}$-singlet operators
  and $f_{e,N-k}$ are the functions of $\mu_{e,n+e}$ and $\mu_{e,n+e+1}$ like \eqref{ef}.
  The equations of motion with respect to $M_{e,k}$ give
    \bea
    \tr \mu_{e,n+e}^{k} - \tr \mu_{e,n+e+1}^{k}
     =     N m_{e} \Lambda_{e}^{2N-1} \delta^{kN} + \frac{N-1}{M_{e}} \Lambda_{e}^{2N-1} \delta^{k,N-1}.
    \eea
  Also, from the $\Phi_{i=n+e}$ equation of motion, we get $\mu_{e-1,n+e} = {}^{t}\mu_{e,n+e}$.

  Let us gauge the diagonal subgroup of $\SU(N)_{n+1}$ and $\SU(N)_{2n+1}$.
  Effectively, this is an $\SU(N)$ gauge theory coupled to the three adjoints $\Phi$,
  $A := \mu_{1,n+1}$ and $B := {}^{t}\mu_{n,2n+1}$ with the superpotential
    \bea
    W
     =     \sum_{k=2}^{N} X_{k} \left( \tr A^{k} - \tr B^{k} - N m \delta^{kN}
         - (N-1) \lambda \delta^{k,N-1} \right) + \tr \Phi (A - B),
    \eea
  where $m= \sum_{e} m_{e} \Lambda_{e}^{2N-1}$, $\lambda = \sum_{e} \frac{\Lambda^{2N-1}_{e}}{M_{e}}$
  and $X_{k}$ are Lagrange multipliers.
  The F-term equation with respect to $\Phi$ gives $A = B$,
  while the F-term equations with respect to $X_{k}$ restrict the terms in the parenthesis to vanish for each $k$.
  Thus the supersymmetry is broken unless $m = \lambda = 0$.

  This can be seen as conditions on residues of the functions $V_{N}$ and $V_{N-1}$ in the curve.
  In general, for each holomorphic one-form $\lambda_i~(i=1,\cdots,g)$ on the base Riemann surface $\CC_{g,n}$, we can define
  meromorphic one-forms $V_k \lambda_i$. The sum of residues of these meromorphic one-forms must vanish
  in any compact Riemann surface. The above field theory constraints on parameters corresponds to
  the constraints on the residues of $V_k$ in the curve.

\paragraph{Pinching of the curve}
  Let us consider the graph with $g$ loops and $n$ $\SU(N)_{e}$ gauge groups with $N_{e}$ flavors ($e=1,\ldots,n$).
  By the decoupling of an $\CN=2$ vector multiplet,
  in some cases the graph is split into two graphs, one with $g_{1}$ loops and $n_{1}$ $\SU(N)_{e_{1}}$ gauge groups
  with $N_{e_{1}}$ flavors ($e_{1} = 1,\ldots,n_{1}$),
  and one with $g_{2}$ loops and $n_{2}$ $\SU(N)_{e_{2}}$ gauge groups
  with $N_{e_{2}}$ flavors ($e_{2} = 1,\ldots,n_{2}$),
  where $g_{1} + g_{2} =g$ and $n_{1} + n_{2} = n$.
  In the other cases the graph is simply reduced to the one with $g-1$ loops
  and the same $\SU(N)_{e}$ gauge groups.

  This can be easily seen from the curve.
  Each $\SU(N)_{e}$ gauge group corresponds to a point $p_e$ on the genus $g$ Riemann surface.
  The function $V_{k}$ has a simple pole at the point $p_{e}$ if $N-N_{e} \leq k \leq N$.
  In the first cases, the Riemann surface is split
  into two Riemann surfaces: one with genus $g_{1}$ and points $p_{e_{1}}$ ($e_{1} = 1,\ldots, n_{1}$)
  and one with genus $g_{2}$ and points $p_{e_{2}}$ ($e_{2} = 1,\ldots, n_{2}$).
  Accordingly, $V_{k}$ becomes two functions: $V_{1k}$ on the former surface
  with simple poles at $p_{e_{1}}$ if $N-N_{e_{1}} \leq k \leq N$,
  and $V_{2k}$ on the latter surface with simple poles at $p_{e_{2}}$ if $N-N_{e_{2}} \leq k \leq N$.
  In the second cases, $V_{k}$ becomes a meromorphic function on the pinched genus $g-1$ surface
  whose singularity at $p_{e}$ is the same as the original $V_{k}$.
  It is important that $V_{k}$ are meromorphic {\it functions} in this analysis.

\section{Conclusion and discussions}
\label{sec:discussion}
 We have studied the low energy dynamics of the $\CN=1$ system consisting of copies of the $T_{N}$ theory and
 $\CN=1$ and $\CN=2$ vector multiplets.
 The $T_{N}$ theory coupled to an $\CN=1$ vector multiplet (with the additional flavor) displays confinement
 in the IR where the effective theory is described by the gauge invariant operators with quantum deformed constraints
 (with the dynamically-generated superpotential).
 The $T_{N}$ theory coupled to two or three $\CN=1$ vector multiplets is in the Coulomb phase with moduli space of vacua,
 whose low energy holomorphic coupling matrix is identified with the period matrix of an $\CN=1$ Seiberg-Witten curve.
 We have also generalized it to $\CN=1$ systems denoted by generalized quiver graphs and determined their curves.
 It should be emphasized that while the system is not a conventional one, we can confirm these results
 by using the methods widely used in the literature: holomorphy and symmetry, 't Hooft anomaly matching, and so on.

 The curve of the $\CN=1$ systems discussed in the previous section is reminiscent
 of the Seiberg-Witten curve of $\CN=2$ gauge theory
 which is understood widely from the M-theoretical point of view \cite{Witten:1997sc,Gaiotto:2009we,Gaiotto:2009hg}.
 So, it would be interesting to construct the $\CN=1$ system by using M5-branes
 or if one prefers, just the 6d $\cN=(2,0)$ theory.
 Let us briefly discuss this here, leaving details to future work.
 Recall that the $T_N$ theory is obtained by putting $N$ M5-branes on a sphere with three maximal punctures.
 A maximal puncture is the intersections of $N$ M5-branes with a particular choice of other branes.
 Many other types of punctures are known, which correspond to the intersections with other combinations of branes.
 The punctures are divided into two main classes, the tame ones and the wild ones.
 Each intersection locally preserves half of the supersymmetry.
 If we decide to use all the punctures to preserve the same half of the supersymmetry,
 we can realize $\cN=2$ supersymmetric theory which can be either tame \cite{Gaiotto:2009we,Nanopoulos:2009xe}
 or wild \cite{Gaiotto:2009hg,Nanopoulos:2010zb,Cecotti:2011rv,Bonelli:2011aa,Xie:2012hs,Gaiotto:2012rg,Kanno:2013vi}.
 We can also make some of the punctures to preserve different half of the supersymmetry,
 realizing $\cN=1$ supersymmetric theory in the end.
 The tame cases have been analyzed before \cite{Maruyoshi:2009uk,Benini:2009mz,Bah:2012dg,Beem:2012yn,Gadde:2013fma}
 and correspond to $\CN=1$ theories which flow to IR superconformal fixed points.
 The $\CN=1$ system analyzed above is the wild version.
 With the brane construction, it would be possible to derive the results in the previous sections geometrically. 

\section*{Acknowledgments}
It is a pleasure to thank Francesco Benini, Yu Nakayama,  Shlomo Razamat 
and Dan Xie for useful comments and helpful discussions. 
The work of K.M. is supported by a JSPS postdoctoral fellowship for research abroad. The work of Y.T. is supported in part by World Premier International Research Center Initiative  (WPI Initiative),  MEXT, Japan through the Institute for the Physics and Mathematics of the Universe, the University of Tokyo. The work of W.Y. is supported in part by the Sherman Fairchild scholarship and by DOE grant DE-FG02-92-ER40701.
The work of K.Y. is supported in part by NSF grant PHY- 0969448.

\appendix
\section{Detailes of Higgsing to the SQCD results}
\label{appsec:ConstraintsClosingPunc}

  When one puncture is closed to the minimal puncture by giving a vev to $\mu_A$ as
  $\vev{\mu_A}=\rho_{\star}( \sigma^+)$, the $T_N$ theory is reduced
  to a bifundamental hyper multiplet ${\cal Q}^{i_B i_C}, \tilde{{\cal Q}}_{i_B i_C}$
  and some Nambu-Goldstone multiplets.
  In this appendix we provide all the details needed to show that the deformed moduli constraints
  in section \ref{sec:deformedresult} and the effective superpotential \ref{subsec:Nc+1} reduce to
  those of $\SU(N)$ SUSY QCD with $N_f=N$ and $N_{f}=N+1$ flavors respectively.

  As in \eqref{eq:GenPuncDecomp},
  the adjoint representation of $\SU(N)$ decomposes as a representation of $\SU(2) \times \U(1)$ as
    \beq
    {\rm adj}
     =     \bigoplus_{j=1}^{N-2} V_{j,0} \oplus V_{0,0} \oplus V_{\frac{N-2}{2},N}
           \oplus V_{\frac{N-2}{2},-N},
    \eeq
  where $V_{j,q}$ is a spin $j$ representation of $\SU(2)$ with $\U(1)$ charge $q$.
  Then the $N^2-1$ generators of $\SU(N)_A$ decomposes into
  $T_{j,m,0}$ ($m=-j,-j+1,\cdots,j$), $T_{\U(1)}$ and $T_{\frac{N-2}{2},m,\pm N}$ ($m=-(N-2)/2,\cdots, (N-2)/2$),
  which satisfy the commutation relations \eqref{comm}.
  The argument below \eqref{comm} leads to
    \bea
    \mu_A
     =     \rho_{\star}( \sigma^+)+\sum_{j=1}^{N-2} T_{j,-j,0} \mu_A^{j,-j,0}+T_{\U(1)} \mu_A^{0,0,0}
         + T_{\frac{N-2}{2},-\frac{N-2}{2},\pm N} \mu_A^{\frac{N-2}{2},-\frac{N-2}{2},\pm N}.
           \label{eq:fixedmuA}
    \eea

  We take the matrix form of $\rho_\star$
  such that the only nonzero components are $ (\rho_\star(\sigma^\pm))^k_{~k \pm 1} \neq 0~(k =1,\cdots, N-2$)
  and $(\rho_\star(\sigma^3))^k_{~k} =(N-2k)/2~~(k=1,\cdots,N-1)$.
  Then the only nonzero components of the matrix $T_{j,-j,0}$ are $(T_{j,-j,0})^{k}_{~k-j}~(k=j+1,\cdots,N-1)$.
  For the matrices $ T_{\frac{N-2}{2},-\frac{N-2}{2},\pm N}$,
  we take $(T_{\frac{N-2}{2},-\frac{N-2}{2}, N})^{N-1}_{~N} =(T_{\frac{N-2}{2},-\frac{N-2}{2}, -N})^{N}_{~1}=1 $
  and other components are zero.
  Finally, $T_{\U(1)}={\rm diag}(1,\cdots,1,-(N-1))$.
  By using these, a matrix representation of (\ref{eq:fixedmuA}) is given as
    \beq
    \mu_A
     \sim  \left( \begin{array}{cccc|c}
           \mu_A^{0,0,0} & (\rho_\star(\sigma^+))^1_{~2}  & 0 & 0&0 \\ 
           \vdots & \ddots &  \ddots & 0 & \vdots \\ 
           \mu_A^{N-3,-(N-3),0} &    & \ddots & (\rho_\star(\sigma^+))^{N-2}_{~N-1}  & 0 \\
           \mu_A^{N-2,-(N-2),0} &\mu_A^{N-3,-(N-3),0}  & \ldots& \mu_A^{0,0,0} & \mu_A^{\frac{N-2}{2},-\frac{N-2}{2}, N}
           \\ \hline
           \mu_A^{\frac{N-2}{2},-\frac{N-2}{2},- N} & 0 & \ldots & 0 & -(N-1)\mu_A^{0,0,0}
           \end{array} \right),
           \label{eq:matrixformA}
    \eeq
  where we have neglected order one coefficients of components containing $\mu_A^{j,-j,0}~(j \geq 1)$.

\paragraph{SQCD with $N_{f}=N$:}
  Now let us study the constraints of SQCD with $N_{f}=N$:
  $\tr \mu_A^k=\tr \mu_B^k~(k=2,\cdots,N-1)$ and $\tr \mu_A^N-\tr \mu_B^N  =N\Lambda^{2N}$ under the above Higgsing.
  One can see that
    \beq
    \tr \mu_A^k \propto \mu_A^{k-1,-(k-1),0}+\cdots, ~~~(k=2,\cdots,N-1) \label{eq:muApower}
    \eeq
  where the ellipsis denotes a sum of products of $\mu_A^{j,-j,0}$'s with $0\leq j<k-1$.
  Thus $\tr \mu_A^k=\tr \mu_B^k~(k=2,\cdots,N-1)$ just determines $\mu_A^{k-1,-(k-1),0}$
  in terms of $\mu_A^{0,0,0}$ and $\mu_B$.
  We do not try to determine the explicit form of them.

Next let us consider the equation
\beq
P_B(-(N-1)\mu_A^{0,0,0})-P_A(-(N-1)\mu_A^{0,0,0})=\Lambda^{2N}. \label{eq:specialxvalue}
\eeq
Relations $\tr \mu_A^k=\tr \mu_B^k~(k=2,\cdots,N-1)$ guarantee that this equation is equivalent to the constraint $\tr \mu_A^N-\tr \mu_B^N  =N\Lambda^{2N}$.
Directly from the definition $P_A(-(N-1)\mu_A^{0,0,0})=\det [-(N-1)\mu_A^{0,0,0}{\bf 1}-\mu_A]$ and \eqref{eq:matrixformA},
we get
\beq
P_A(-(N-1)\mu_A^{0,0,0})=-C(\rho) \mu_A^{\frac{N-2}{2},-\frac{N-2}{2}, N}\mu_A^{\frac{N-2}{2},-\frac{N-2}{2}, -N},
\eeq
where $C(\rho)$ is a product of components of $\rho_\star(\sigma^+)$ defined as
\beq
C(\rho)=\prod_{k=1}^{N-2} (\rho_\star(\sigma^+))^k_{~k+1}.
\eeq
Then we obtain
\beq
\det (-(N-1)\mu_A^{0,0,0} {\bf 1}-\mu_B)+C(\rho) \mu_A^{\frac{N-2}{2},-\frac{N-2}{2}, N}\mu_A^{\frac{N-2}{2},-\frac{N-2}{2}, -N}=\Lambda^{2N}.
\label{eq:redconstraint}
\eeq
Now we see that equation (\ref{eq:redconstraint}) is indeed the deformed moduli constraint in SQCD $\det \CM-\CB_+ \CB_-=\Lambda^{2N}$
(up to a normalization of $\Lambda^{2N}$) if we assume the following identification;
\beq
\hat{\CM} \equiv \CM-\frac{{\bf 1}}{N} \tr \CM=\mu_B,~~\tr \CM=N(N-1) \mu_A^{0,0,0},~~\CB_\pm=c_\pm \mu_A^{\frac{N-2}{2},-\frac{N-2}{2}, \pm N},\label{eq:idTNSQCD}
\eeq
where $c_\pm$ are constants satisfying $c_+ c_- =(-1)^{N-1}C(\rho)$.

  Let us see this is the case by considering the operators $Q^{i_A i_B i_C}$ and $Q_{i_A i_B i_C}$.
  Under the above Higgsing, we expect the following identification of the bifundamental fields
  ${\cal Q}^{i_B i_C}$ and $\tilde{{\cal Q}}_{i_B i_C}$;
\beq
{\cal Q}^{i_B i_C}=c'_+Q^{1,i_B i_C},~~~\tilde{\cal Q}_{i_B i_C}=c'_-Q_{N-1,i_B i_C}, \label{eq:bifundamentalID}
\eeq
where $c'_\pm$ are constants.
The other components of $Q^{i_A i_B i_C}$ and $Q_{i_A i_B i_C}$ are obtained as follows.
Using \eqref{eq:CR2}, we obtain
\beq
(\mu^k_B)^{i_B}_{~j_B} {\cal Q}^{ j_B i_C}=c'_+(\mu^k_A)^{1}_{~j_A}Q^{j_A i_B i_C} \propto Q^{k+1, i_B i_C}+\cdots~~~(k=1,\cdots,N-2).
\eeq
This equation gives $Q^{k, i_B i_C}~(k=2,\cdots,N-1)$ in terms of $\mu_B$ and ${\cal Q}$. $Q_{k, i_B i_C}~(k=1,\cdots,N-2)$ are determined in a similar way.
We will discuss how to obtain $Q^{N,i_B i_C}$ and $Q_{N,i_B i_C}$ later.

The meson is given as $\CM^{i_B}_{~j_B}={\cal Q}^{i_B i_C}\tilde{{\cal Q}}_{j_B i_C}$. Using \eqref{eq:CR3},
we get
\beq
\CM^{i_B}_{~j_B}&=
c'_+ c'_- \sum_{l=0}^N \sum_{m=0}^{N-l-1} v_l (\mu_A^{N-l-1-m})^{1}_{~N-1} (\mu_B^m)^{i_B}_{~j_B} \nonumber \\
&=c'_+ c'_- C(\rho)\left(\mu_B+(N-1)\mu_A^{0,0,0} {\bf 1} \right)^{i_B}_{~j_B},
\eeq
where we have used $(\mu_A^{k})^{1}_{~N-1}=0$ for $(k=1,\cdots,N-3)$, $(\mu_A^{N-2})^{1}_{~N-1}=C(\rho)$ and
$(\mu_A^{N-1})^{1}_{~N-1}=(N-1)C(\rho)\mu_A^{0,0,0}$. By setting $c'_+ c'_- = C(\rho)^{-1}$, we precisely get the first two equations of
(\ref{eq:idTNSQCD}).

The baryons are given as $\CB_+=\det {\cal Q}$ and $\CB_-=\det \tilde{\cal Q}$. Before considering them,
we first determine $Q^{N,i_B i_C}$ and $Q_{N,i_B i_C}$, which has $\U(1)$ charges $-(N-1)$ and $N-1$ respectively.
We use the constraints \eqref{eq:CR4} and \eqref{eq:CR4'} for $i_{A,1}=\cdots=i_{A,N-1}=1$ and
$i_{A,1}=\cdots=i_{A,N-1}=N-1$ respectively. We get
\beq
&\frac{1}{(N-1)!}{\cal Q}^{i_{B,1} i_{C,1}}{\cal Q}^{ i_{B,2} i_{C,2}}\cdots {\cal Q}^{ i_{B,N-1} i_{C,N-1}}
\epsilon_{ i_{B,1}i_{B,2}\cdots i_{B,N-1} i_B}\epsilon_{ i_{C,1}i_{C,2}\cdots i_{C,N-1} i_C} \nonumber \\
&=(c'_+)^{N-1} Q_{i_A i_B i_C} (\mu_A^0)^1_{~1}(\mu_A)^{1}_{~2} (\mu^2_A)^{1}_{~3}\cdots (\mu^{N-2}_A)^{1}_{~N-1}
\epsilon^{1,2,  \cdots,   N-1, i_A } \nonumber \\
&= (c'_+)^{N-1} C_+(\rho)Q_{N, i_B i_C} \label{eq:QNcomp1}
\eeq
and similarly
\beq
&\frac{1}{(N-1)!}\tilde{\cal Q}_{i_{B,1} i_{C,1}} \tilde{\cal Q}_{ i_{B,2} i_{C,2}}\cdots \tilde{\cal Q}_{ i_{B,N-1} i_{C,N-1}}
\epsilon^{ i_{B,1}i_{B,2}\cdots i_{B,N-1} i_B}\epsilon^{ i_{C,1}i_{C,2}\cdots i_{C,N-1} i_C} \nonumber \\
&=(-1)^{N-1} (c'_-)^{N-1} C_-(\rho)Q^{N, i_B i_C} \label{eq:QNcomp2}
\eeq
where we have defined
\beq
C_+(\rho)=\prod_{k=1}^{N-2} \left(\rho_\star(\sigma^+)^k_{~k+1} \right)^{N-1-k},~~~~~
C_-(\rho)=\prod_{k=1}^{N-2} \left(\rho_\star(\sigma^+)^k_{~k+1} \right)^{k}.
\eeq
These equations determine $Q^{N,i_B i_C}$ and $Q_{N,i_B i_C}$ in terms of ${\cal Q}^{i_B i_C}$ and $\tilde{\cal Q}_{i_B i_C}$.
The baryons can be obtained by multiplying ${\cal Q}^{i_B i_C}$ to (\ref{eq:QNcomp1}) and
$\tilde{\cal Q}_{i_B i_C}$ to (\ref{eq:QNcomp2}).
Using \eqref{eq:CR3}, we obtain
\beq
Q^{1,i_B i_C}Q_{N, i_B i_C}&=NC(\rho)\mu_A^{\frac{N-2}{2},-\frac{N-2}{2}, N} ,  \\
Q^{N, i_B i_C} Q_{N-1, i_B i_C}&=NC(\rho)\mu_A^{\frac{N-2}{2},-\frac{N-2}{2}, - N}.
\eeq
Then, the baryons are given as
\beq
\CB_+&= (c'_+)^{N} C_+ (\rho) C(\rho) \mu_A^{\frac{N-2}{2},-\frac{N-2}{2},  N}. \\
\CB_-&=(-1)^{(N-1)} (c'_-)^{N} C_- (\rho) C(\rho)\mu_A^{\frac{N-2}{2},-\frac{N-2}{2}, - N}.
\eeq
This is consistent with (\ref{eq:idTNSQCD}) if the coefficients satisfy the relation
\beq
(-1)^{N-1} (c'_+ c'_-)^{N}(C_+ (\rho) C_-(\rho)) C(\rho)^2=(-1)^{N-1}C(\rho).
\eeq
Using $c'_+ c'_-=C(\rho)^{-1}$ and $C_+ (\rho) C_-(\rho)=C(\rho)^{N-1}$, we can see that this is really the case.

\paragraph{SQCD with $N_{f} = N+1$:}
  Here let us see the derivation of \eqref{eq:superpiece2} which is needed to get the effective superpotential
  of SQCD with $N_{f} = N+1$.
  By using the equations of motion of massive fields \eqref{eq:massiveeq2},
  the last two terms of the superpotential \eqref{superpotentialNc+1} is
  \beq
  &B_{i_A i_B} \tilde{B}^{i_A j_B} (\mu_B)^{i_B}_{~j_B}
  -B_{i_A i_B} \tilde{B}^{j_A i_B} (\mu_A)^{i_A}_{~j_A} \nonumber \\[0.3cm]
   = &B_{N i_B} \tilde{B}^{N i_B} (\mu_B)^{i_B}_{~j_B}
    +B_{1 i_B} \tilde{B}^{1 j_B} (\mu_B)^{i_B}_{~j_B}
     -  B_{i_A i_B} \tilde{B}^{N i_B} (\mu_A)^{i_A}_{~N}
   -  B_{i_A i_B} \tilde{B}^{1 i_B} (\mu_A)^{i_A}_{~1}  \nonumber \\[0.3cm]
   =&B_{N i_B} \tilde{B}^{N i_B} \CM^{i_B}_{~j_B}
  - B_{N i_B} \tilde{B}^{1 i_B} \mu_{A}^{\frac{N-2}{2},-\frac{N-2}{2},-N}
    -  B_{N-1 i_B} \tilde{B}^{N i_B} \mu_{A}^{\frac{N-2}{2},-\frac{N-2}{2},N}  \nonumber \\
  & +\Big(B_{1 i_B}  \CM^{i_B}_{~j_B}-\sum_{i_A=1}^{N-1}  B_{i_A j_B}  \hat{\mu}^{i_A}_{~1} \Big) \tilde{B}^{1 j_B},
      \label{eq:Bsuperdecompose}
      \eeq
  where we have defined
  $\hat{\mu}=\mu_A+(N-1)\mu_A^{0,0,0} {\bf 1}$.

 Let us determine the last term of \eqref{eq:Bsuperdecompose} in terms of $B_{N-1,i_B}$.
\eqref{eq:massiveeq2} gives us
 \beq
 B_{i_A}=(\rho^{i_A}_{~i_A+1})^{-1} \sum_{i_A<j_A \leq N-1} B_{j_A} (\delta^{j_A}_{~i_A+1} \CM - \hat{\mu}^{j_A}_{~i_A+1})
 \eeq
  where we have omitted the indices of $\SU(N)_B$ by considering $\CM$ and $B_{i_A}$ as a matrix and a vector respectively,
  and we have used abbreviation $\rho=\rho(\sigma^+)$. Repeatedly using this equation, we obtain
  \beq
  &\sum_{0<i_1 \leq N-1} B_{i_1} (\delta^{i_1}_{~1} \CM - \hat{\mu}^{i_1}_{~1}) \nonumber \\
  =&B_{N-1}\sum_{a=0}^{N-2}\sum_{0<i_1<\cdots < i_a < N-1}(\delta^{N-1}_{~i_a+1} \CM - \hat{\mu}^{N-1}_{~i_a+1})\cdot (\rho^{i_a}_{~i_a+1})^{-1} \nonumber \\
&~~~~~  \cdot (\delta^{i_a}_{~i_{a-1}+1} \CM - \hat{\mu}^{i_a}_{~i_{a-1}+1})\cdot (\rho^{i_{a-1}}_{~i_{a-1}+1})^{-1}  \cdot \cdots \cdot
(\rho^{i_{1}}_{~i_1+1})^{-1} \cdot (\delta^{i_1}_{~1} \CM - \hat{\mu}^{i_1}_{~1}). \label{eq:B1toBN}
  \eeq
  By a careful inspection of the matrix \eqref{eq:matrixformA}, we can see that \eqref{eq:B1toBN} is summarized as
  \beq
 C(\rho)^{-1} B_{N-1} P_{\hat{\mu}'}(\CM),
  \eeq
  where $\hat{\mu}'$ is the $N-1 \times N-1$ matrix $\hat{\mu}^{i}_{~j}$ with $1 \leq i,j \leq N-1$, and $P_{\hat{\mu}'}(x)=\det(x {\bf 1}-\hat{\mu}')$
  is the characteristic polynomial of $\hat{\mu}'$. This is related to $P_{\hat{\mu}}(x)=\det(x {\bf 1}-\hat{\mu})$ as
  \beq
  P_{\hat{\mu}}(x)=xP_{\hat{\mu}'}(x)-C(\rho)\mu_{A}^{\frac{N-2}{2},-\frac{N-2}{2},N} \mu_{A}^{\frac{N-2}{2},-\frac{N-2}{2},-N}.
  \eeq
  Because of the relation $\tr \mu_A^k=\tr \mu_B^k~(k=2,\cdots,N-1)$, we have
  \beq
   P_{\hat{\mu}}(x)=P_{\CM}(x)+(x~{\rm independent~term})
  \eeq
  where $P_\CM(x)=\det (x {\bf 1}-\CM)$. Combining the above equations, we get
  \beq
  P_{\hat{\mu}'}(x)=x^{-1} (P_\CM(x)-(-1)^{N} \det \CM).
  \eeq
  Therefore, \eqref{eq:B1toBN} is finally reduced to
  \beq
  (-1)^{N-1}C(\rho)^{-1} B_{N-1}\CM^{-1} \det \CM ,
  \eeq
  where we have used $P_\CM(\CM)=0$.
  We identify
    \beq
    &\CB_{i_B}
    =   - (c'_+)^{N-1} C_+(\rho) B_{N i_B}, ~~~
    \tilde{\CB}^{i_B}
     =     (-1)^{N} (c'_-)^{N-1} C_-(\rho) B^{N i_B}, \nonumber \\
    &\CM^{i_B}_{~N+1}
    =     c'_+\tilde{B}^{1 i_B}, ~~~~~~~~~~~~~~~~
    \CM^{N+1}_{~i_B}
     =     c'_- B_{N-1 i_B}. \label{eq:Nf=N+1ID}
    \eeq
  Using $c_+=(c'_+)^{N} C_+ (\rho) C(\rho)$ and $c_-=(-1)^{N-1}  (c'_-)^{N} C_- (\rho) C(\rho)$,
  the superpotential terms \eqref{eq:Bsuperdecompose} is now rewritten as \eqref{eq:superpiece2}.

\bibliographystyle{ytphys}
\bibliography{ref}

\end{document}